\documentclass[preprint,9pt]{elsarticle}

\usepackage{amssymb}
\usepackage{amsmath}
\usepackage{bm}
\usepackage{tikz}
\usepackage{enumitem}
\usepackage{subcaption}
\usepackage{float}
\usetikzlibrary{arrows.meta, positioning}
\usepackage{hyperref}
\usepackage{mathtools}
\usepackage{makecell}

\usepackage{url}

\journal{Applied Mathematics Letters}

\begin{document}

\begin{frontmatter}



\title{Uncertain data assimilation for urban wind flow simulations with OpenLB-UQ} 


\author[SCC,LBRG,IANM]{Mingliang Zhong}
\author[LBRG,MVM]{Dennis Teutscher}
\author[LBRG,IANM]{Adrian Kummerl\"{a}nder}
\author[LBRG,IANM,MVM]{Mathias J. Krause}
\author[SCC,IANM]{Martin Frank}
\author[LBRG,IANM]{Stephan Simonis}

\affiliation[SCC]{organization={Scientific Computing Center, Karlsruhe Institute of Technology},
            addressline={Eggenstein-Leopoldshafen}, 
            city={Karlsruhe},
            postcode={76344}, 
            state={Baden-Württemberg},
            country={Germany}}
            
\affiliation[IANM]{organization={Institute for Applied and Numerical Mathematics, Karlsruhe Institute of Technology},
            addressline={Englerstr. 2}, 
            city={Karlsruhe},
            postcode={76131}, 
            state={Baden-Württemberg},
            country={Germany}}

\affiliation[MVM]{organization={Institute of Mechanical Process Engineering and Mechanics},
            addressline={Straße am Forum 8}, 
            city={Karlsruhe},
            postcode={76131}, 
            state={Baden-Württemberg},
            country={Germany}}

\affiliation[LBRG]{organization={Lattice Boltzmann Research Group, Karlsruhe Institute of Technology},
            addressline={Englerstr. 2}, 
            city={Karlsruhe},
            postcode={76131}, 
            state={Baden-Württemberg},
            country={Germany}}

\begin{abstract}
Accurate prediction of urban wind flow is essential for urban planning, pedestrian safety, and environmental management. 
Yet, it remains challenging due to uncertain boundary conditions and the high cost of conventional CFD simulations. 
This paper presents the use of the modular and efficient uncertainty quantification (UQ) framework OpenLB-UQ for urban wind flow simulations. 
We specifically use the lattice Boltzmann method (LBM) coupled with a stochastic collocation (SC) approach based on generalized polynomial chaos (gPC). 
The framework introduces a relative-error noise model for inflow wind speeds based on real measurements. 
The model is propagated through a non-intrusive SC LBM pipeline using sparse-grid quadrature. 
Key quantities of interest, including mean flow fields, standard deviations, and vertical profiles with confidence intervals, are efficiently computed without altering the underlying deterministic solver. 
We demonstrate this on a real urban scenario, highlighting how uncertainty localizes in complex flow regions such as wakes and shear layers. 
The results show that the SC LBM approach provides accurate, uncertainty-aware predictions with significant computational efficiency, making OpenLB-UQ a practical tool for real-time urban wind analysis.
\end{abstract}




\end{frontmatter}


\section{Introduction}

Assessing wind comfort~\cite{jacob2018wind}, structural loading~\cite{zhang2021cfd}, and pollutant dispersion~\cite{teutscher2025digital} in cities demands flow predictions that faithfully represent urban geometries and are able to quantify uncertainty.
High-fidelity CFD approaches, such as direct numerical simulation (DNS) and large-eddy simulation (LES), can resolve key flow structures around buildings, but even a single simulation is computationally costly.
When combined with uncertainty quantification (UQ), which typically requires an ensemble of simulations, the cost becomes prohibitive.
Practical urban-flow UQ simulations thus call for combining computationally efficient CFD solvers with equally efficient UQ methods that can estimate quantities of interest (QoIs)—such as the mean and variance of flow variables—with accuracy competitive with Monte Carlo under smooth QoIs, but at significantly reduced cost. 

While UQ has been widely studied in the context of Reynolds-averaged Navier--Stokes (RANS)-based urban flow simulations, its integration with LES is still emerging. 
Recent advances have started to address both the stochastic variability inherent to LES and the systematic propagation of uncertain boundary conditions or parameters.
Lumet \textit{et al.}~\cite{Lumet2024InternalVar} proposed a methodology to quantify the internal variability of LES for pollutant dispersion in urban canopies, highlighting the impact of finite sampling and turbulence chaos on predictive uncertainty. 
Building upon this work, Lumet \textit{et al.}~\cite{Lumet2025Surrogate} introduced an uncertainty-aware reduced-order model trained on a 200-member LES ensemble. 
More recently, a surrogate-assisted ensemble data assimilation method has been proposed to further reduce prediction uncertainty in
LES-based pollutant dispersion studies~\cite{Lumet2025DA}. 
Parallel developments include work on LES ensemble strategies for variable inflow conditions. 
Keskinen and Hellsten~\cite{Keskinen2025Ensembles} analyzed ensembles of urban LES with changing wind directions, providing guidance on the required sample size and representativeness for UQ applications. 
Overall, these contributions mark an important step toward the systematic integration of UQ and LES for urban wind flow and pollutant dispersion. Nevertheless, the literature remains sparse compared to RANS-based UQ, and LES-based UQ studies are still at an early stage. 
This suggests considerable potential for further methodological developments and applications in this area. 
To the best of our knowledge, so far no study on combining efficient LES of real urban environments with sophisticated UQ-methods in a single simulation framework exists. 

LES based on the lattice Boltzmann method (LBM) is well suited for efficiently simulating urban wind flow at larger scales: its explicit, local update rules yield excellent parallel scalability and high computational efficiency, while naturally accommodating complex urban geometries.
When coupled with non-intrusive stochastic collocation (SC) using generalized polynomial chaos (gPC)~\cite{Xiu2005, Xiu2002, Nobile2008Sparse, Eldred2008}, LBM can propagate input uncertainty with Monte-Carlo-level accuracy for smooth QoIs using far fewer solver evaluations \cite{zhong2025openlbuq}.
This combined efficiency conceptually enables large-scale UQ studies of urban aerodynamics. 

In many practical applications, the dominant uncertainty arises from boundary conditions based on measurements, such as time-varying inflow wind speed and direction~\cite{garcia2014quantifying}.
A rigorous treatment of measurement-driven inputs should account for both additive and multiplicative (relative) sensor errors.
Over much of a sensor’s operating range, gain-type (multiplicative) uncertainty makes the standard deviation approximately a constant fraction of the signal, while at low speeds, an absolute error term dominates.
For example, wind speed measurements may specify an uncertainty of $\pm 0.5~\mathrm{m\,s^{-1}}$ for speeds $\le 5~\mathrm{m\,s^{-1}}$, transitioning to $\pm 10\%$ for higher speeds~\cite{WMO8}.

In this work, we present an uncertain data assimilation workflow based on OpenLB-UQ~\cite{zhong2025openlbuq}, which directly injects measurement uncertainty into boundary data and propagates it through an SC LBM simulation of the urban wind flow around an isolated real building geometry in the city of Reutlingen (Germany, \(48.4914^\circ\mathrm{N}, 9.2043^\circ\mathrm{E}\)).  
The resulting stochastic boundary condition is mapped to a dense quadrature grid in the gPC space, and OpenLB-UQ is used to compute an ensemble of LBM solutions at the collocation nodes.
From these, we recover spatio-temporal statistics (means, variances, and confidence bands) of wind speed and velocity magnitude throughout the urban domain, including diagnostics tailored to urban applications (e.g., wake regions, shear layers, and nominally open-channel corridors). 

This paper is structured as follows. 
In Section~\ref{sec:methodology}, we formulate the stochastic problem statement, introduce the deterministic LBM-based solver as a non-intrusive core, as well as the SC-gPC methodology to approximate the system response, and present the complete workflow. 
In Section~\ref{sec:setup}, we describe the numerical setup in OpenLB-UQ and the inclusion of uncertainty in the measurement data. 
Section~\ref{sec:results} summarizes the uncertain data-assimilated simulation results. 
Section~\ref{sec:conclusion} draws conclusions and suggests future research directions.

\section{Methodology} \label{sec:methodology}

\subsection{Problem statement and uncertainty channels}

We consider the unsteady incompressible Navier--Stokes equations with uncertainty, formulated as
\begin{align}
    \partial_t \bm{u}(\bm{Z}) + (\bm{u}(\bm{Z})\cdot \bm{\nabla}) \bm{u}(\bm{Z}) - \nu \bm{\nabla}^2 \bm{u}(\bm{Z}) &= -\frac{1}{\rho} \bm{\nabla} p(\bm{Z}),
    && \text{in } \mathcal{X} \times \mathcal{I} \times \mathcal{Z}, \label{eq:stochNSEmom} \\
    \bm{\nabla} \cdot \bm{u}(\bm{Z}) &= 0, 
    && \text{in } \mathcal{X} \times \mathcal{I} \times \mathcal{Z}, \label{eq:stochNSEdiv}
\end{align}
where the \( \bm{u} \colon \mathcal{X} \times \mathcal{I} \times \mathcal{Z} \to \mathbb{R}^{d_x} \) is the stochastic velocity field, \( p\colon \mathcal{X} \times \mathcal{I} \times \mathcal{Z} \to \mathbb{R}\) is the stochastic pressure, and \( \nu >0 \) is the viscosity.
The uncertainty is modeled by a vector of independent random variables \( \bm{Z} = (Z_1, \ldots, Z_{d_Z}) \in \mathcal{Z}\) defined on a probability space \( (\Omega, \mathcal{F}, \mathbb{P}) \).

In addition, we prescribe deterministic initial conditions and stochastic boundary conditions to \eqref{eq:stochNSEmom} and \eqref{eq:stochNSEdiv}, which are described below (see Section~\ref{sec:detLBM}), to form a stochastic initial boundary value problem that models uncertain wind flow in an urban geometry.
In this work, we focus on uncertainty in the inlet velocity due to measurement errors. 
Specifically, we model the uncertainty in the measured wind speed  and direction with a respective relative perturbation \(\zeta_{1}\) and \(\zeta_{2}\), resulting in a logarithmic wind profile 
\begin{align}  \label{eq:mult_noise}
\hat{\bm{y}}(\bm{x}, t, \bm{Z}), \quad \text{where } \bm{Z} \coloneqq  \bm{\zeta} = (\zeta_{1}, \zeta_{2})^{\mathrm{T}} \sim \mathcal{N}(1, 0.1^{2})\times\mathcal{N}(0, 6^{2})
\end{align}
that is further described below in Section~\ref{sec:setup} and realized as a convective boundary condition at the horizontal domain boundaries. 
This stochastic boundary condition introduces temporal but spatially coherent uncertainty into the domain.

\subsection{Deterministic lattice Boltzmann method (non-intrusive core)} \label{sec:detLBM}

We approximate the deterministic version of~\eqref{eq:stochNSEmom} and \eqref{eq:stochNSEdiv} using a recursive regularized LBM based on a third-order expanded equilibrium function and combined with a Smagorinsky--Lilly subgrid scale (SGS) model. 
An advanced version of this collision model, including hybridization and homogenization, is further described in \cite{teutscher2025digital}. 
The following quantities are thus assumed as filtered variables and not further denoted as such. 
\begin{figure}[ht!]
\centering
\begin{tikzpicture}[>=latex]
 \centering
    \draw[densely dashed] (-2,0,0) -- (0,0,0);
    \draw[densely dashed] (-2,0,0) -- (-2,2,0);
    \draw[densely dashed] (-2,0,0) -- (-2,0,2);
    \draw[densely dashed] (-2,0,2) -- (0,0,2);
    \draw[densely dashed] (-2,2,0) -- (0,2,0);
    \draw[densely dashed] (-2,0,2) -- (-2,2,2);
    \draw[densely dashed] (-2,2,0) -- (-2,2,2);
    \draw[densely dashed] (-2,2,2) -- (0,2,2);
    \draw[densely dashed] (0,0,0) -- (0,0,2);
    \draw[densely dashed] (0,0,0) -- (0,2,0);
    \draw[densely dashed] (0,0,2) -- (0,2,2);
    \draw[densely dashed] (0,2,2) -- (0,2,0);
 	\draw[densely dashed] (-2,0,1) -- (0,0,1); 
	\draw[densely dashed] (-2,0,1) -- (-2,2,1); 
	\draw[densely dashed] (-2,2,1) -- (0,2,1); 
	\draw[densely dashed] (0,2,1) -- (0,0,1); 
	\draw[densely dashed] (-1,0,0) -- (-1,0,2); 
	\draw[densely dashed] (-1,0,2) -- (-1,2,2); 
	\draw[densely dashed] (-1,2,2) -- (-1,2,0); 
	\draw[densely dashed] (-1,2,0) -- (-1,0,0); 
	\draw[densely dashed] (-2,1,0) -- (0,1,0); 
	\draw[densely dashed] (0,1,0) -- (0,1,2); 
	\draw[densely dashed] (0,1,2) -- (-2,1,2); 
	\draw[densely dashed] (-2,1,2) -- (-2,1,0);   
    \draw[->,thick](-1,1,1) -- (-1,1,0);
    \draw[->,thick](-1,1,1) -- (-2,1,0);
    \draw[->,thick](-1,1,1) -- (-1,2,0);
    \draw[->,thick](-1,1,1) -- (-1,0,0);
    \draw[->,thick](-1,1,1) -- (0,1,0);
    \draw[thick,fill=cyan](-1,1,0) circle(2pt);     
	\draw[thick,fill=green](-2,1,0) circle(1.5pt);
  	\draw[thick,fill=green](-1,2,0) circle(1.5pt);
  	\draw[thick,fill=green](-1,0,0) circle(1.5pt);
	\draw[thick,fill=green](0,1,0) circle(1.5pt);
    \draw[->,thick](-1,1,1) -- (-1,2,1);
    \draw[->,thick](-1,1,1) -- (0,1,1);
    \draw[->,thick](-1,1,1) -- (-2,1,1);
    \draw[->,thick](-1,1,1) -- (-1,0,1);
    \draw[->,thick](-1,1,1) -- (-2,0,1);
    \draw[->,thick](-1,1,1) -- (0,0,1);
    \draw[->,thick](-1,1,1) -- (-2,2,1);
    \draw[->,thick](-1,1,1) -- (0,2,1);
	\draw[thick,fill=cyan](-1,2,1) circle(2pt);
    \draw[thick,fill=cyan](0,1,1) circle(2pt);
    \draw[thick,fill=cyan](-1,0,1) circle(2pt);
    \draw[thick,fill=cyan](-2,1,1) circle(2pt);
    \draw[thick,fill=green](-2,0,1) circle(1.5pt);     
  	\draw[thick,fill=green](0,0,1) circle(1.5pt);
    \draw[thick,fill=green](-2,2,1) circle(1.5pt);     
  	\draw[thick,fill=green](0,2,1) circle(1.5pt);
	\draw[->,thick](-1,1,1) -- (-1,1,2);
    \draw[->,thick](-1,1,1) -- (-1,0,2);
	\draw[->,thick](-1,1,1) -- (-2,1,2);
	\draw[->,thick](-1,1,1) -- (0,1,2);
	\draw[->,thick](-1,1,1) -- (-1,2,2);
    \draw[thick,fill=cyan](-1,1,2) circle(2pt);
	\draw[thick,fill=green](-1,0,2) circle(1.5pt);
	\draw[thick,fill=green](-2,1,2) circle(1.5pt);
	\draw[thick,fill=green](0,1,2) circle(1.5pt);
	\draw[thick,fill=green](-1,2,2) circle(1.5pt);
    \draw[thick,fill=orange](-1,1,1) circle(3pt);
  \end{tikzpicture}
\hspace{2em}
\begin{tikzpicture}
	\draw[->,thick] (0,0,0) -- (.5,0,0);
    \node[anchor=north] at (.5,0,0) {$x$};
    \draw[->,thick] (0,0,0) -- (0,.5,0);
    \node[anchor=east] at (0,.5,0) {$y$};
	\draw[->,thick] (0,0,0) -- (0,0,.5);
	\node[anchor=north west] at (0,0,.5) {$z$};
\end{tikzpicture}
  \caption{A schematic illustration of the discrete velocity set \(D3Q19\). 
  Coloring refers to energy shells: orange, cyan, green denote zeroth, first, second order, respectively. 
  Figure from \cite{simonis2023pde}.}
\label{fig:lattice}
\end{figure}
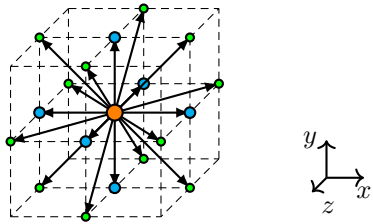

The evolution equation for the (filtered) particle distribution function \( \bm{f} = \{f_i\}_{i=0,1,\ldots,q-1} \in \mathbb{R}^{q} \) is given by
\begin{align}\label{eq:deterministicLBE}
    f_{i} (\bm{x}+\bm{c}_i \triangle t, t+\triangle t) 
    = 
    f_{i}^{\mathrm{eq}} (\bm{x}, t) + \left( 1 - \frac{1}{\tau_{\mathrm{eff}}(\bm{x},t)} \right)f_{i}^{(1)}(\bm{x}, t), 
    \quad \text{in } \mathcal{X}_{\triangle x} \times \mathcal{I}_{\triangle t},
\end{align}
for \(i=0, 1, \ldots, q-1\), where \( \bm{c}_i \) are the discrete lattice velocities of the $D3Q19$ model and \((\bm{x}, t)\) are grid nodes in the Cartesian discrete space-time cylinder \(\mathcal{X}_{\triangle x} \times \mathcal{I}_{\triangle t}\). 
Moreover, \( \tau_{\mathrm{eff}} \) is the space-time adaptive, effective relaxation time of the SGS model, given by  
\begin{equation}\label{eq:tauEff}
    \tau_\mathrm{eff}(\bm{x},t) = \frac{\nu_\mathrm{eff}(\bm{x},t)}{c_{\mathrm{s}}^2} \frac{\triangle t}{\triangle x^2} + \frac{1}{2},
\end{equation}
with \( \nu_\mathrm{eff} = \nu + \nu_\mathrm{turb} \) representing the combined molecular (\(\nu\)) and turbulent viscosity 
\begin{align}
    \nu_\mathrm{turb}(\bm{x},t)  = \left(C_{\mathrm{S}} \triangle_{\bm{x}}\right)^2 \left|\mathbf{S}(\bm{x},t)\right|, \label{eq:turbVisc}
\end{align}  
where $C_{\mathrm{S}} = 0.25$ is the Smagorinsky constant, $\triangle_{\bm{x}} = \triangle x$ is the filter width, and $\mathbf{S}$ is the (filtered) strain rate tensor
\begin{equation}
    S_{\alpha\beta} = \frac{1}{2} \left( \frac{\partial u_{\alpha}}{\partial x_\beta} + \frac{\partial u_{\beta}}{\partial x_\alpha} \right).
\end{equation} 
Here, \( c_{\mathrm{s}} \) is the lattice speed of sound, and we use \( c_{\mathrm{s}}^2 = 1/3 \) in lattice units.
The regularization in \eqref{eq:deterministicLBE} is based on the non-equilibrium function $f_{i}^{(1)} = f_{i} - f_{i}^{(0)}$ that is expanded as
\begin{equation}
  f_{i}^{(1)}(\bm{x},t) = \omega_i \sum_{n=0}^{N=3} \frac{1}{c_{\mathrm{s}}^{2n} n!} \mathbf{H}_i^{(n)} : \mathbf{a}_1^{(n)}(\bm{x},t) ,
\end{equation}
where \(\omega_i\) are the lattice weights and we denote by \( \mathbf{H}_i^{(n)} \) the \(n\)th order Hermite polynomial with the \(i\)th discrete velocity \(\bm{c}_{i}\) as an argument. 
The Hermite coefficient for the non-equilibrium is defined as
\begin{equation}\label{eq:hermiteCoeff}
    \mathbf{a}_1^{(n)}(\bm{x},t)=\sum_{i=0}^{q-1}\mathbf{H}_i^{(n)}f_i^{(1)}(\bm{x},t) .
\end{equation}
Note that according to \cite{JACOB18}, we use Hermite polynomials that have correct orthogonality for \(D3Q19\) only. 
The equilibrium distribution \( f_i^{\mathrm{eq}} (\bm{f}) \) is computed by an expansion using Hermite polynomials and Hermite equilibrium coefficients (\eqref{eq:hermiteCoeff} for \(f^{\mathrm{eq}}_{i}\)) up to order three, i.e.
\begin{equation}
  f_i^{\mathrm{eq}}(\bm{x},t) = \omega_i \rho(\bm{x},t)\left( 1 + \frac{\bm{c}_i \cdot (\bm{u}(\bm{x},t))}{c_{\mathrm{s}}^2}
    + \frac{\mathbf{H}_i^{(2)} : \mathbf{a}_{\mathrm{eq}}^{(2)}(\bm{x},t)}{2 c_{\mathrm{s}}^4 \rho(\bm{x},t)}
    + \frac{\mathbf{H}_i^{(3)} : \mathbf{a}_{\mathrm{eq}}^{(3)}(\bm{x},t)}{2 c_{\mathrm{s}}^6\rho(\bm{x},t)}
  \right),
\end{equation}
where the zeroth-order moment is denoted by \(\rho=\sum_{i=0}^{q-1} f_{i}\) and the first-order local velocity moment is \( \bm{u} = (1/\rho)\sum_{i=0}^{q-1}\bm{c}_{i}f_{i}\). 

The computational domain \(\mathcal{X}_{\triangle x}\) is configured as a disk with thickness $H$ and radius $R$, where the building geometry is centered on the ground. 
The boundary conditions are handled as follows:
\begin{itemize}\setlength{\itemsep}{0.0em}
    \item \textbf{Horizontal boundaries (mantle of cylinder):} The surrounding wind velocity \( \hat{\bm{y}}(\bm{x}, t,\bm{Z}) \) is prescribed via a (circular) local velocity boundary condition~\cite{olbPaper2021}, where both the magnitude and the wind flow direction are treated as uncertain quantities (see Section~\ref{sec:setup}).
    \item \textbf{Building walls and ground (bottom of cylinder):} A no-slip condition is enforced with the classical bounce-back method.
    \item \textbf{Sky (top of cylinder):} A free-slip condition is used by enforcing a zero normal velocity and a zero normal gradient of the tangential velocity.
\end{itemize}

The populations are initialized to equilibrium values, and the horizontal boundary conditions are increased to the first target values over time to stabilize the simulation. 
For a smooth change in the magnitude and direction of horizontal velocity boundaries, an interpolation time interval is used, and the boundary velocity is gradually transformed to the subsequent magnitude and direction. 

\subsection{Generalized polynomial chaos and stochastic collocation} \label{sec:scgpc}

To propagate the input uncertainty through the LBM solver, we use the gPC expansion method implemented in OpenLB-UQ \cite{zhong2025openlbuq}. 
Let \( \bm{Z} \in \mathbb{R}^{d_Z} \) denote the stochastic input vector with joint density \( h(\bm{Z}) \). 
The model output \( \mathcal{M}(\bm{Z}) \) is approximated as
\begin{equation} \label{eq:gPC}
    \mathcal{M}(\bm{Z}) \approx \mathcal{M}^{(N)}(\bm{Z}) 
    = \sum_{\alpha=0}^{N} \hat{\mathcal{M}}_{\alpha}\,\Phi_{\alpha}(\bm{Z}),
\end{equation}
where \( \{ \Phi_\alpha \} \) is an orthonormal polynomial basis (Hermite polynomials for Gaussian input), and \( \hat{\mathcal{M}}_\alpha \) are the expansion coefficients, computed via projection
\begin{equation} \label{eq:coef_discrete}
    \hat{\mathcal{M}}_\alpha 
    \approx \frac{1}{\gamma_\alpha} \sum_{j=1}^{N_q} a_j\,\mathcal{M}(\bm{Z}^{(j)})\,\Phi_\alpha(\bm{Z}^{(j)}),
\end{equation}
where \( \gamma_\alpha = \mathbb{E}[\Phi_{\alpha}^2] \), and \( \{ \bm{Z}^{(j)}, a_j \}_{j=1}^N \) are the quadrature nodes and weights, respectively. 
For scalar or field-valued QoIs, the statistical moments are computed using
\begin{equation}
    \mathbb{E}[Q] \approx \sum_{j=1}^{N_{q}} a_j Q(\bm{Z}^{(j)}), \qquad
    \text{Var}[Q] \approx \sum_{j=1}^{N_{q}} a_j \left(Q(\bm{Z}^{(j)}) - \mathbb{E}[Q]\right)^2.
    \label{eq:quad_moments}
\end{equation}

\subsection{Stochastic collocation lattice Boltzmann coupling and workflow} \label{sec:algo}

The SC LBM is non-intrusive: 
the deterministic LBM solver runs independently at each quadrature node, and statistical analysis is performed afterward \cite{zhong2025openlbuq}. 
The workflow is as follows:
\begin{enumerate}\setlength{\itemsep}{0.0em}
    \item \textbf{Select input measurement data.}
    Hourly wind speed \( U_H(t) \) and direction \(\Theta(t)\) recorded at the Reutlingen station (Germany, \(48.4914^\circ\mathrm{N},\, 9.2043^\circ\mathrm{E}\)), from 2024-11-07 20:00 to 2024-11-09 19:00 (CET).

    \item \textbf{Define the input model.}  
    Specify uncertain parameters, e.g., inlet velocity magnitude 
    \( \zeta_1 \sim \mathcal{N}(1, 0.1^2) \) and inflow direction 
    \( \zeta_2 \sim \mathcal{N}(0, 6^2) \).  
    Select an appropriate polynomial basis (e.g., Hermite) for the expansion.

    \item \textbf{Generate the quadrature rule.}  
    Construct collocation nodes \( \{ \bm{Z}^{(j)} \} \) and weights \( \{ a_j \} \) 
    using a dense tensor-product Gauss--Hermite quadrature.

    \item \textbf{Apply boundary conditions.}  
    For each node \( j \), compute the inlet velocity uncertainty
    \( \hat{\bm{y}}(\bm{x}, t, \bm{Z}^{(j)}) \) via Eq.~\eqref{eq:mult_noise}.

    \item \textbf{Run deterministic simulations.}  
    For each \( \bm{Z}^{(j)} \), solve the LBM problem including hourly uncertain measurement data-assimilation during the simulation. Obtain the sample velocity
    \( \bm{u}(\bm{x}, t , \bm{Z}^{(j)}) \).

    \item \textbf{Extract QoIs.}  
    Examples include:
    \begin{itemize}\setlength{\itemsep}{0.0em}
        \item \emph{Domain-wide fields:} single-sample, time-dependent and time-averaged velocity magnitudes at a given time (e.g., Fig.~\ref{fig:single_sample10_Q-crit_local} and Fig.~\ref{fig:single_sample60_Q-crit_local}).
        \item \emph{Local probes:} vertical velocity profiles at different probe locations (e.g., Fig.~\ref{fig:urban_probes_profiles}).
    \end{itemize}

    \item \textbf{Compute statistics.}  
    Evaluate mean (e.g., Fig.~\ref{fig:meanStdLocal_Q} and Fig.~\ref{fig:meanStdAverage_Q}), variance (e.g., Fig.~\ref{fig:urban_mean_std_contour}), and confidence intervals (e.g., Fig.~\ref{fig:urban_probes_profiles}) using 
    Eqs.~\eqref{eq:quad_moments} or the gPC expansion coefficients.
\end{enumerate}

\section{Simulation case setup and parameter configuration}\label{sec:setup}


To represent input uncertainty in the inflow boundary condition, both the reference wind speed and direction are modeled with artificial measurement perturbation, respectively 
\begin{align}
  \widehat{U}_H(t, \zeta_{1}) &= U_H(t)\zeta_1 , 
  & \zeta_1 &\sim \mathcal{N}\!\left(1,\,0.1^2\right), \notag \\
  \widehat{\Theta}(t, \zeta_{2}) &= \Theta(t) + \zeta_2,
  & \zeta_2 &\sim \mathcal{N}\!\left(0,\,6^2\right), \notag
\end{align}
where \(U_H(t)\) and \(\Theta(t)\) are the recorded wind speed and direction (the latter is given in radians).
In summary, we prescribe a logarithmic wind profile at horizontal boundaries 
\begin{align}
\hat{\bm{y}}(\bm{x}, t, \bm{Z}) = \hat{\bm{y}}(\bm{x}, t, \bm{\zeta}) =  \dfrac{\widehat{U}_{H} (t, \zeta_{1}) }{\ln\left( \dfrac{H + z_0}{z_0} \right)} \ln\left( \dfrac{h  + z_0}{z_0} \right)
\begin{pmatrix}
\cos\bigl(\widehat{\Theta}(t, \zeta_{2})\bigr)  \\
\sin\bigl(\widehat{\Theta}(t, \zeta_{2})\bigr) \\
0
\end{pmatrix} , 
\end{align}
where the roughness length of the surface is \(z_0 = 0.1\)[m], \(H\)[m] is the reference height and \(h = |z|\) is the vertical coordinate in meters above ground, where \(\bm{x}= (x,y,z)^{\mathrm{T}} \in \mathbb{R}^{3}\).

\begin{figure}[ht!]
  \centering
  \begin{subfigure}[t]{\textwidth}
    \centering
    \includegraphics[width=0.75\textwidth]{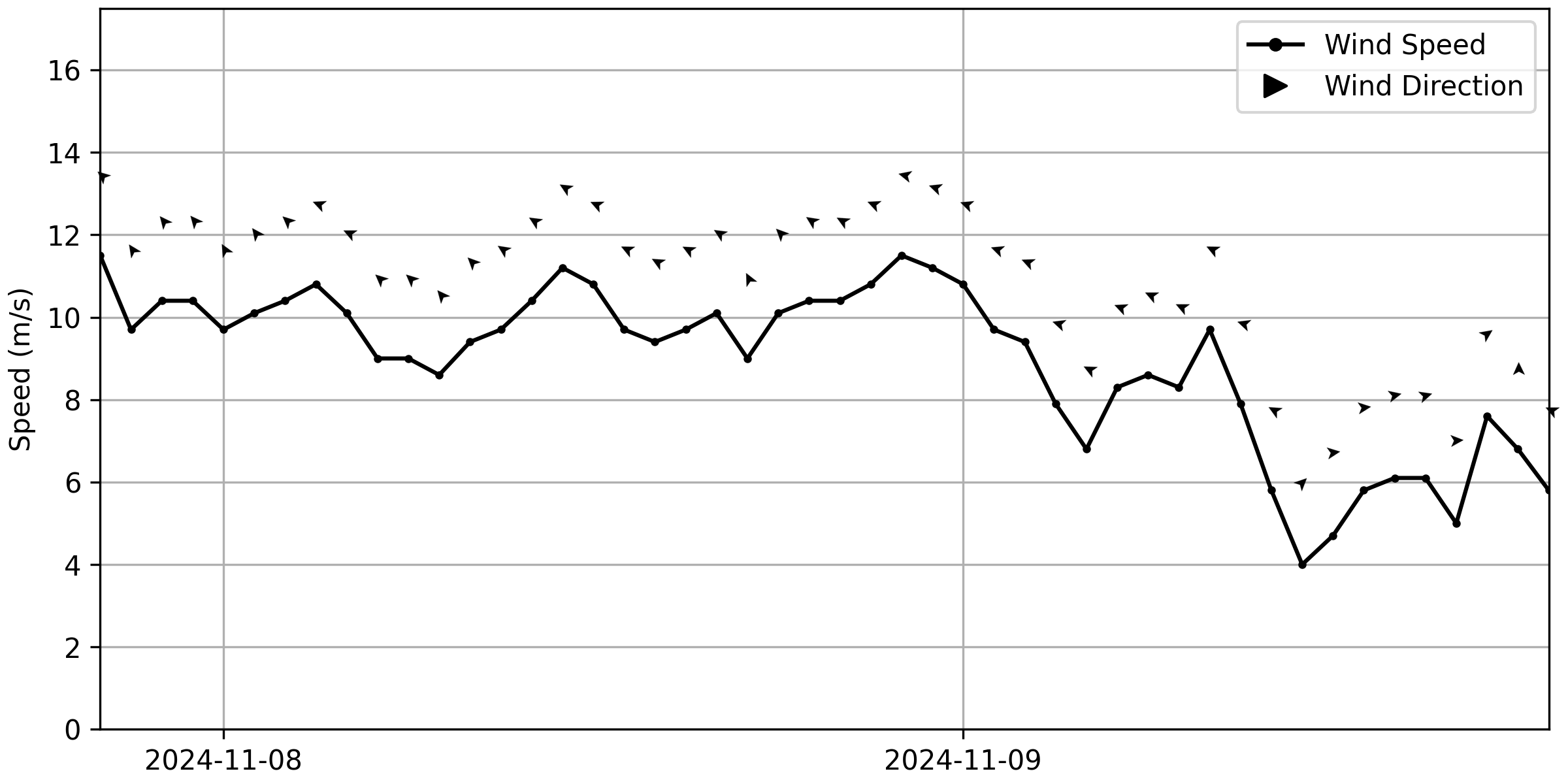}
    \caption{Wind speed and wind direction measurements}
    \label{fig:wind_speed_direction_sub}
    \vspace{.5em}
  \end{subfigure}
  \begin{subfigure}[t]{\textwidth}
    \centering
    \includegraphics[width=.75\textwidth]{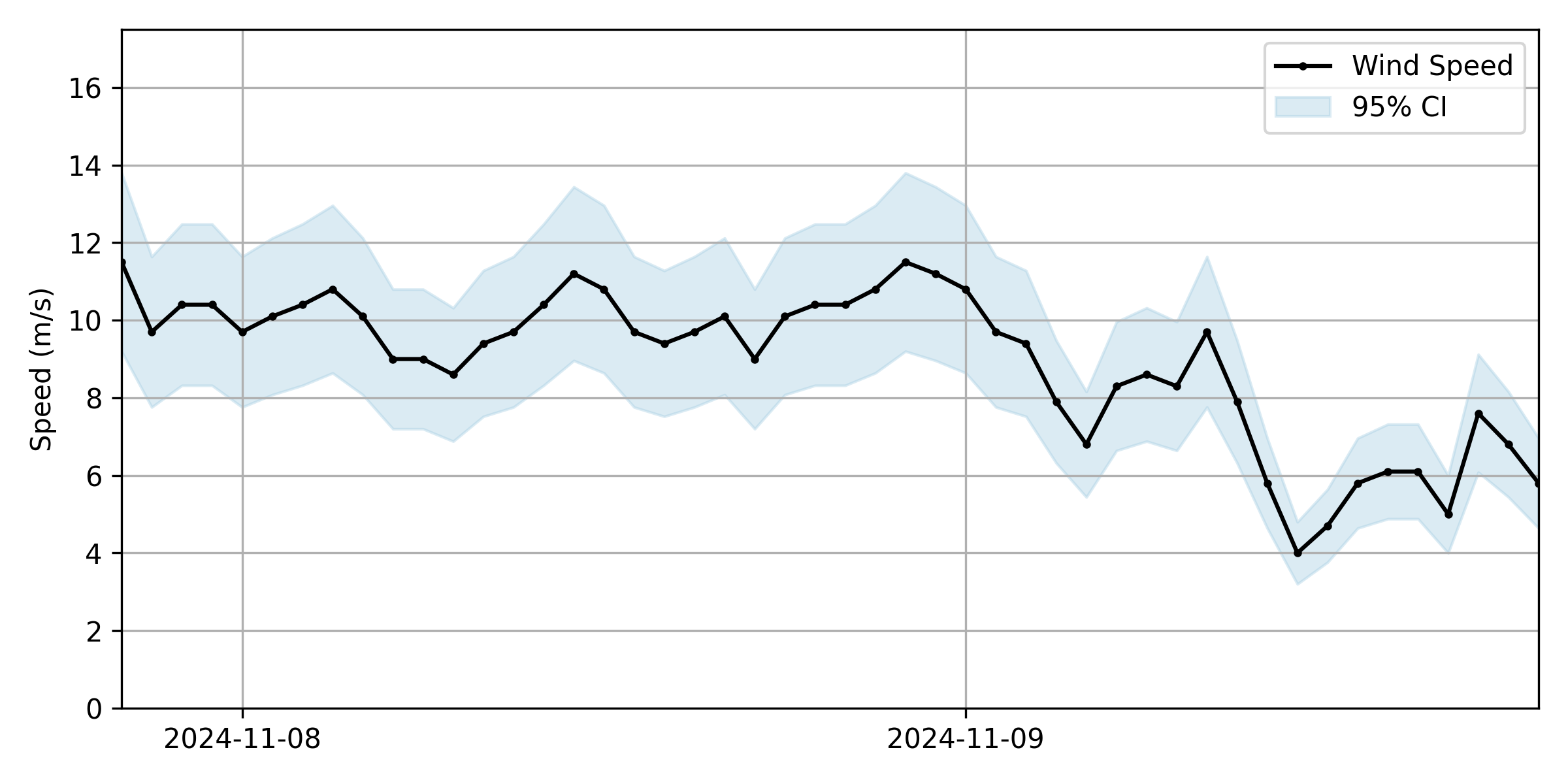}
    \caption{Wind speed measurements with 95\% confidence interval}
    \label{fig:wind_speed_uncertainty_sub}
  \end{subfigure}
    \caption{
        Hourly reference wind speed \(U_H(t)\) recorded at the Reutlingen station (Germany, \(48.4914^\circ\mathrm{N},\, 9.2043^\circ\mathrm{E}\)) from 2024-11-07 20:00 to 2024-11-09 19:00 (local time, CET), obtained from \url{https://meteostat.net/}.
        Subfigure~(a) shows wind speed and direction; subfigure~(b) displays wind speed with the shaded \(95\%\) confidence interval.
    }
  \label{fig:wind_speed_overview}
\end{figure}

Figure~\ref{fig:wind_speed_direction_sub} shows the hourly reference wind speed \(U_H(t)\) together with wind direction, recorded at the Reutlingen station (Germany, \(48.4914^\circ\mathrm{N},\, 9.2043^\circ\mathrm{E}\)) from 2024-11-07 20:00 to 2024-11-09 19:00 (local time, CET).
Figure~\ref{fig:wind_speed_uncertainty_sub} displays the same wind speed data along with the shaded \(95\%\) confidence interval (CI) band.
Our simulations use the 48-hour measurement shown in Figure~\ref{fig:wind_speed_overview}.

The simulation was performed using a dense-grid quadrature scheme with a polynomial order of \(N = 5\) and 11 quadrature points per stochastic dimension, resulting in a total of \(N_q = 121\) collocation points (i.e.\ samples).
The physical and numerical parameters of the case are summarized in Table~\ref{tab:phys_num_params}. 
The simulation domain is configured as a circular disk region (with radius \(R=270\) [m] and height \(H=40\) [m]. 
Two buildings from the city of Reutlingen are placed in the center of the domain.

\begin{table}[ht!]
\centering
\footnotesize
\caption{Physical and numerical parameters.}
\label{tab:phys_num_params}
\begin{tabular}{cccccc} 
\hline
$\triangle x$  & $\triangle t$ & $Ma$ & $Re$ & $T$ & \#cells \\
\hline
1 [m] & 0.00111 [s] & $0.034641$ & $1.8\times 10^{6}$ & 48 [h] & $10.27 \times 10^{6}$ \\
\hline
\end{tabular}
\end{table}
All simulations were executed on the HoreKa supercomputer on CPU-only compute nodes equipped with two \textit{Intel Xeon Platinum 8368} processors, providing 76 physical cores (152 hardware threads) per node at 2.40~GHz.
At the job level, 42 samples were launched concurrently across 168 nodes, with each sample run in parallel on 304 MPI ranks. The wall-clock time per simulation was \(5.4 \times 10^{4}\,\mathrm{s}\) (about 15.0 hours), corresponding to approximately 4560 CPU core-hours per sample.
In total, 121 samples were completed within a turnaround time of about 2.1 days (Fri Aug 22 22:51:52 2025 to Mon Aug 25 01:29:35 2025), yielding a total computational cost of about \(5.5 \times 10^{5}\) CPU core-hours, followed by an additional \(3.6 \times 10^{3}\,\mathrm{s}\) (1.0 hour) for postprocessing.
Theoretically, with 484 dedicated nodes, the sampling stage alone could be finished in \(5.4 \times 10^{4}\,\mathrm{s}\) (15 hours).

In the SC LBM framework, uncertainty is propagated by assigning each collocation node \(j\) a realization of the random multipliers \(\zeta_1^{(j)}\) and \(\zeta_2^{(j)}\), drawn from \(\mathcal{N}(1,\,0.1^2)\) and \(\mathcal{N}(0,\,6^2)\).

\section{Results}\label{sec:results}

\begin{figure}[ht!]
    \centering
    \includegraphics[trim={0 6cm 0 95cm},clip,width=0.5\textwidth]{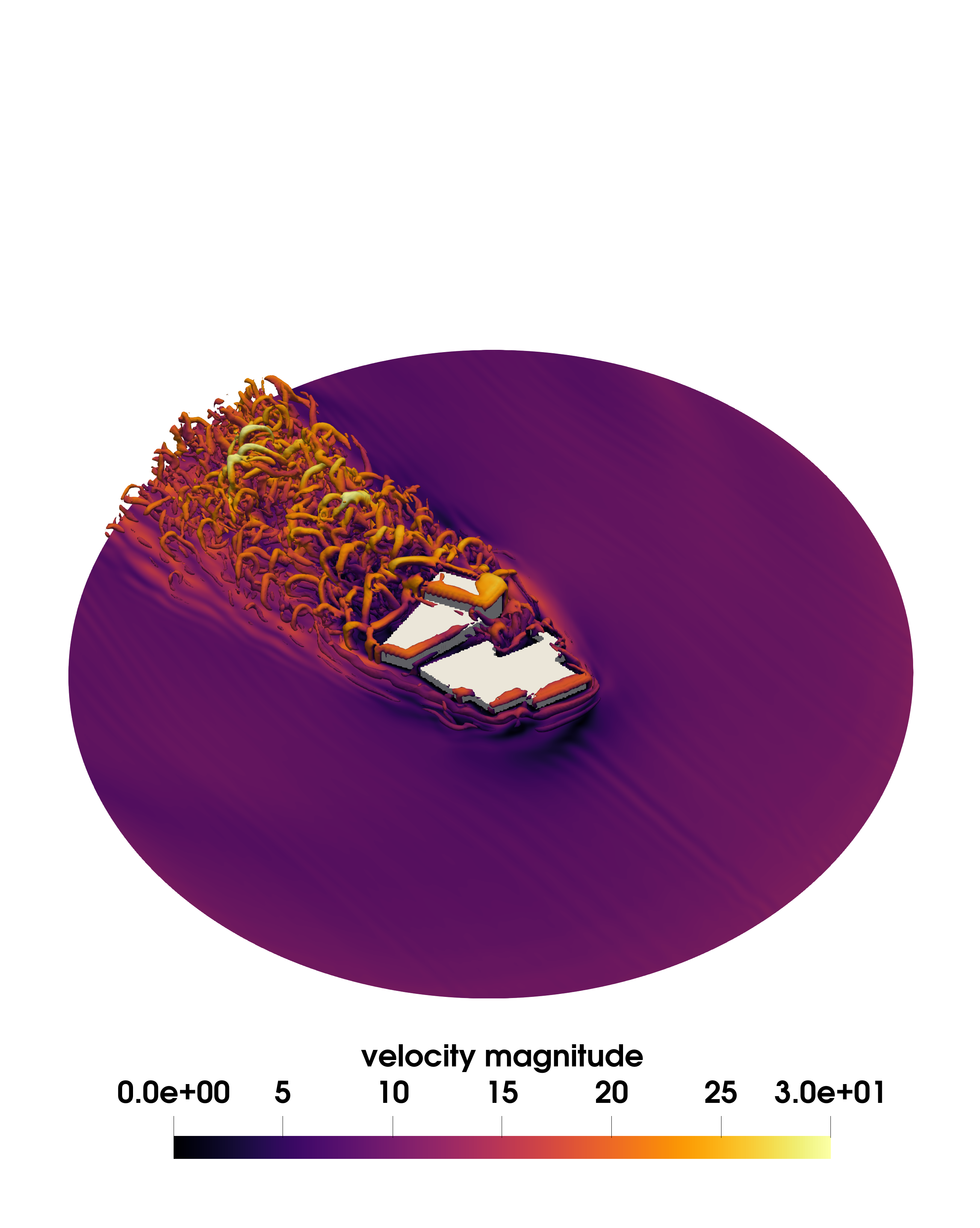}\\
    \subfloat[2024-11-07 22:00:00 CET]{\includegraphics[trim={0 20cm 0 25cm},clip,width=0.32\textwidth]{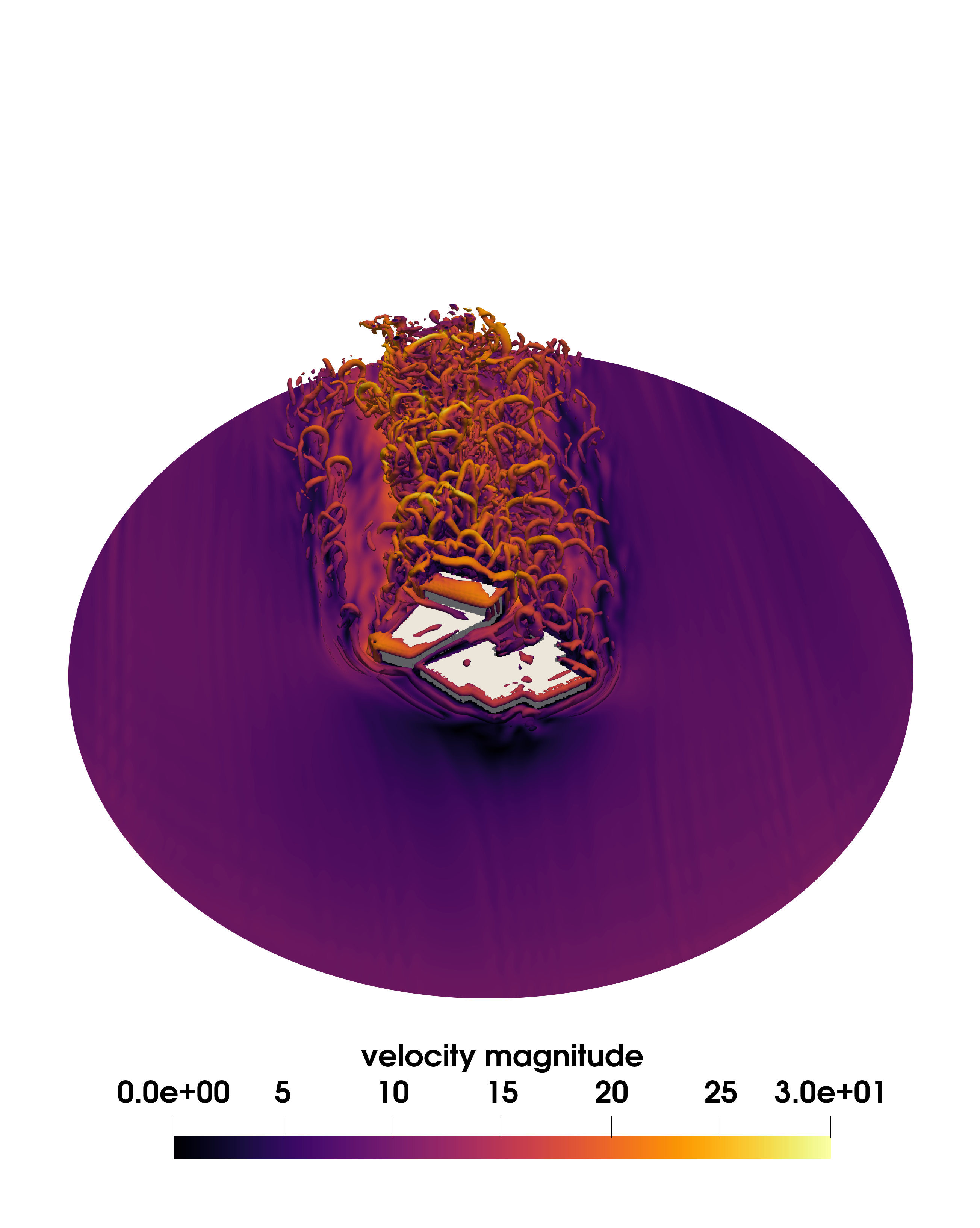}}
    \subfloat[2024-11-08 10:00:00 CET]{\includegraphics[trim={0 20cm 0 25cm},clip,width=0.32\textwidth]{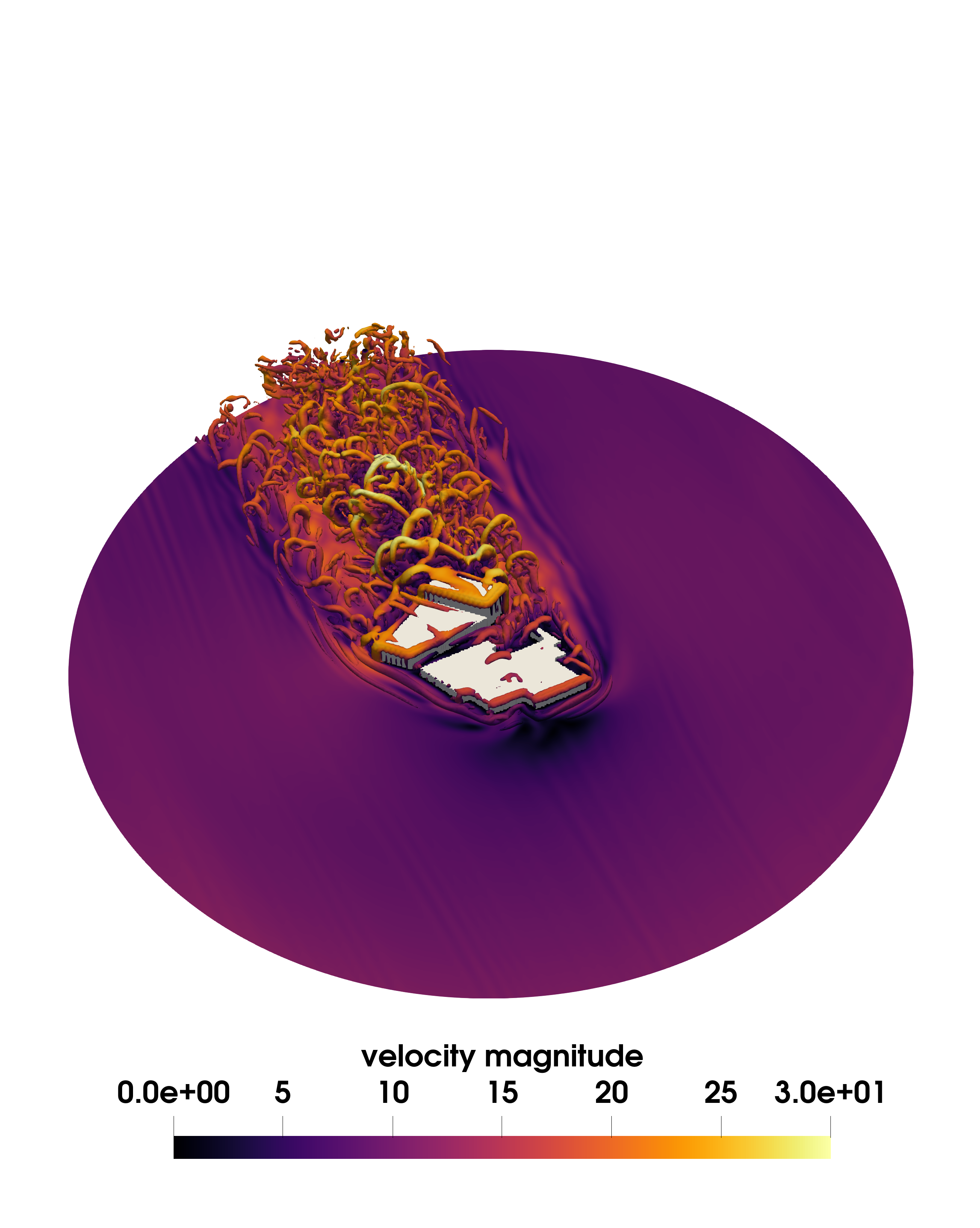}}
    \subfloat[2024-11-08 21:00:00 CET]{\includegraphics[trim={0 20cm 0 25cm},clip,width=0.32\textwidth]{fig/singleSample/magmaLocalVelSingleSample10.0027.png}}\\
    \includegraphics[trim={0 6cm 0 95cm},clip,width=0.5\textwidth]{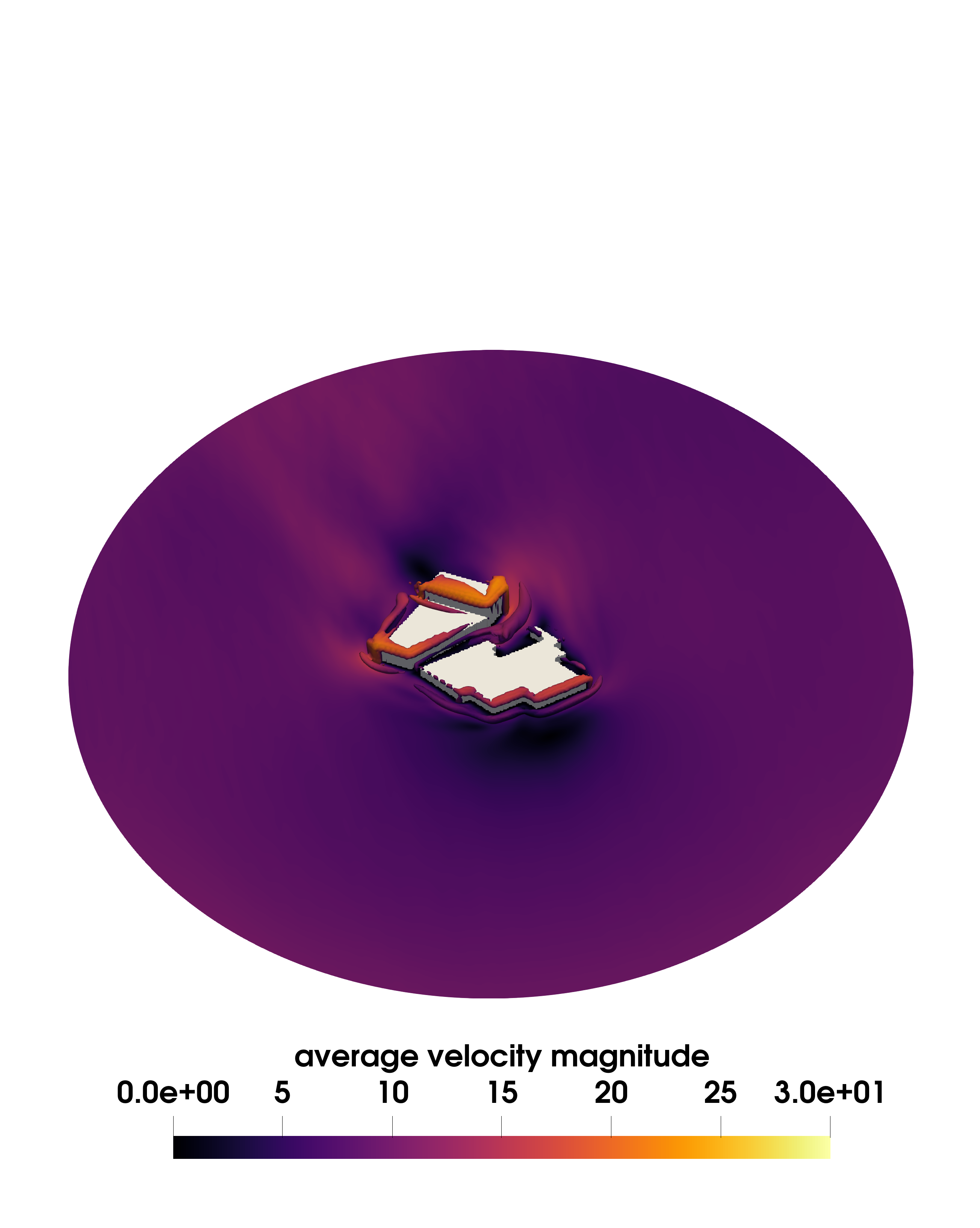}\\
    \subfloat[2024-11-07 22:00:00 CET]{\includegraphics[trim={0 20cm 0 25cm},clip,width=0.32\textwidth]{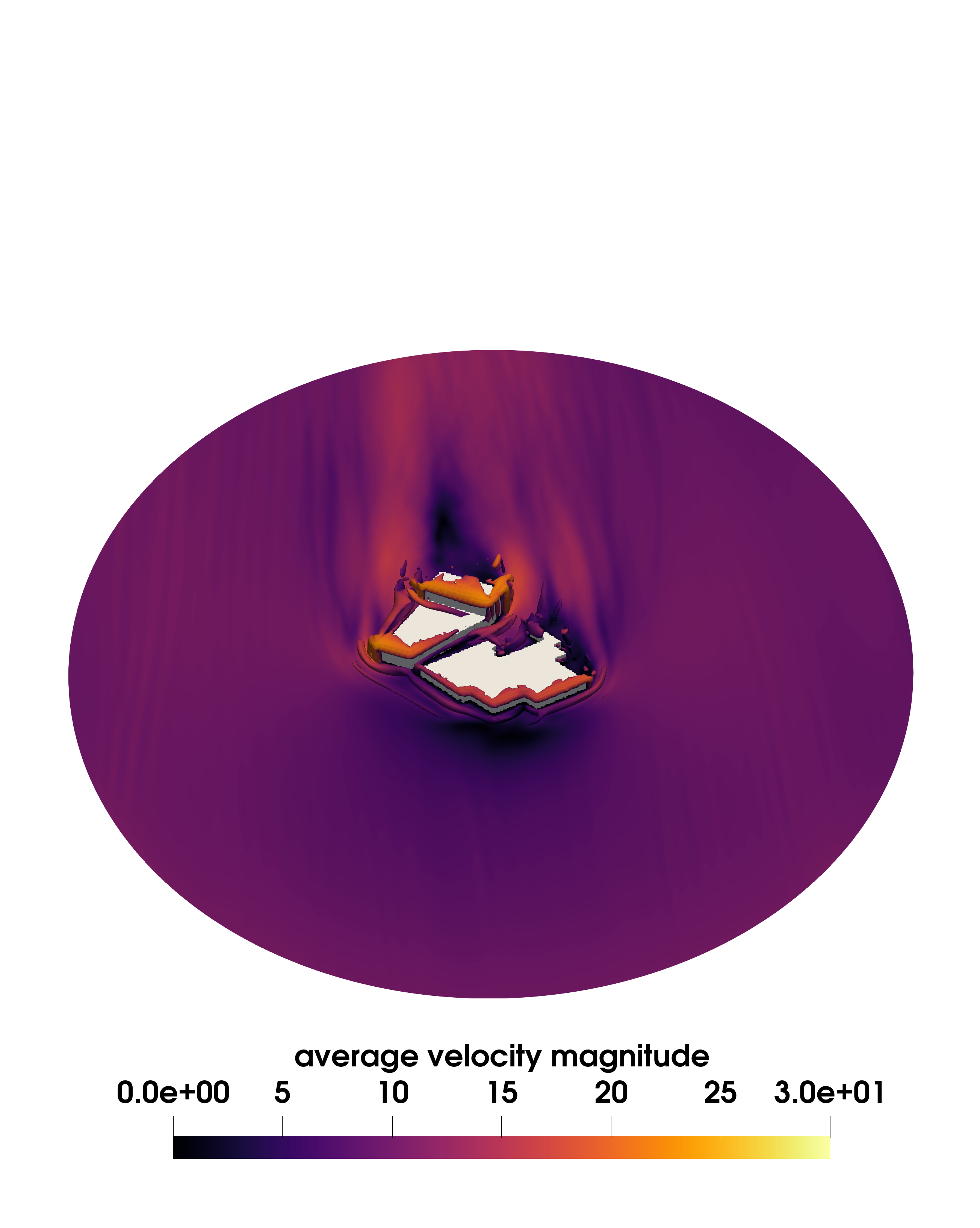}}  
    \subfloat[2024-11-08 10:00:00 CET]{\includegraphics[trim={0 20cm 0 25cm},clip,width=0.32\textwidth]{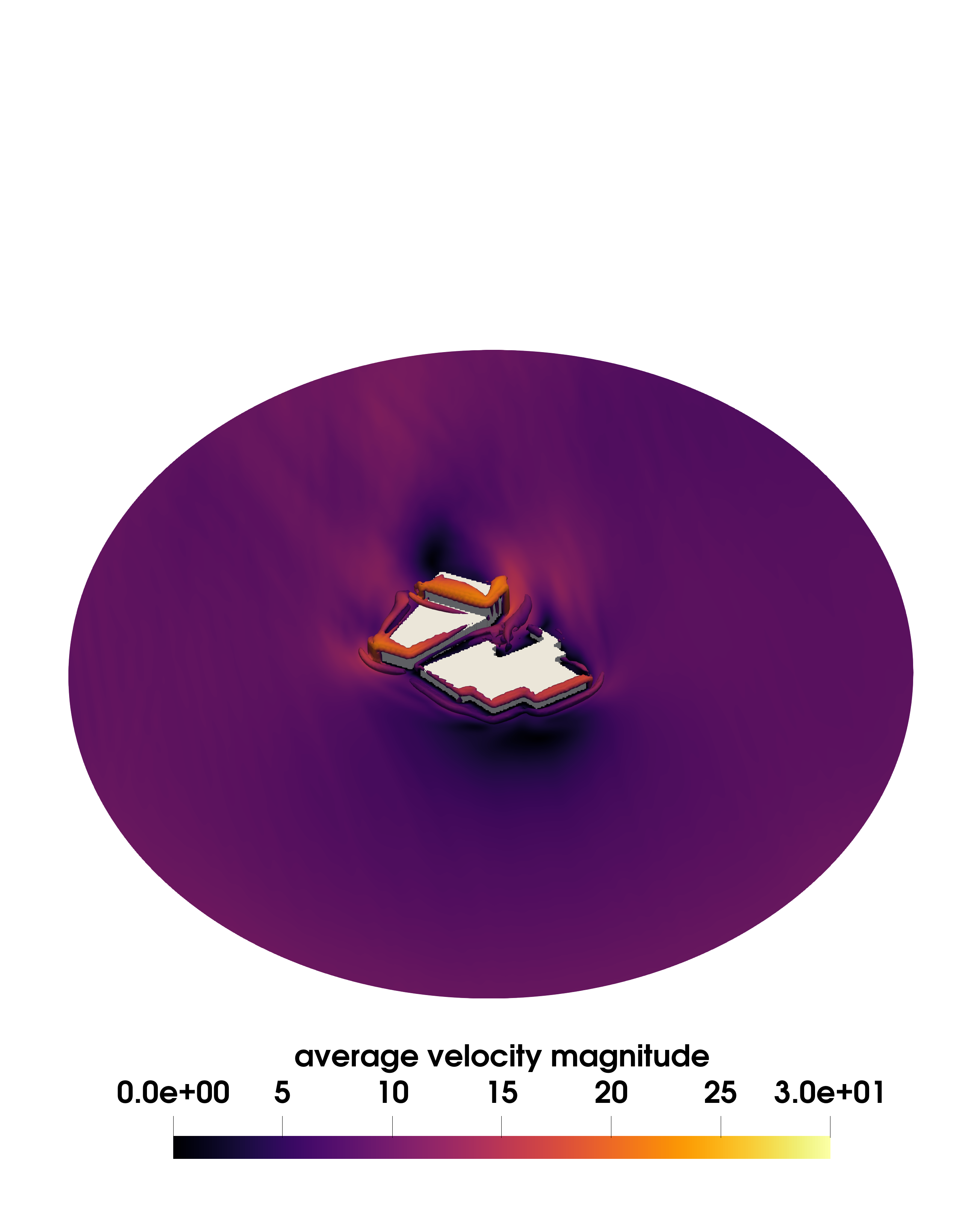}}
    \subfloat[2024-11-08 21:00:00 CET]{\includegraphics[trim={0 20cm 0 25cm},clip,width=0.32\textwidth]{fig/singleSample/magmaAveraVelSingleSample10.0027.png}}
    \caption{
     Simulated (SC LBM) single sample (number 10 out of 121) time-dependent velocity field magnitude at (a) 2024-11-07 22:00:00 CET, (b) 2024-11-08 10:00:00 CET, (c) 2024-11-09 08:00:00, and time-averaged velocity field magnitude at (d) 2024-11-07 22:00:00 CET, (e) 2024-11-08 10:00:00 CET, (f) 2024-11-09 08:00:00, respectively. 
     Isocontours are colored in velocity magnitude and represent the \(Q\)-criterion at \(Q=1\), of the local-in-time results (a,b,c) and the time-averaged time averaged results (d,e,f), respectively.
    }
    \label{fig:single_sample10_Q-crit_local}
\end{figure}

\begin{figure}[ht!]
    \centering
    \includegraphics[trim={0 6cm 0 95cm},clip,width=0.5\textwidth]{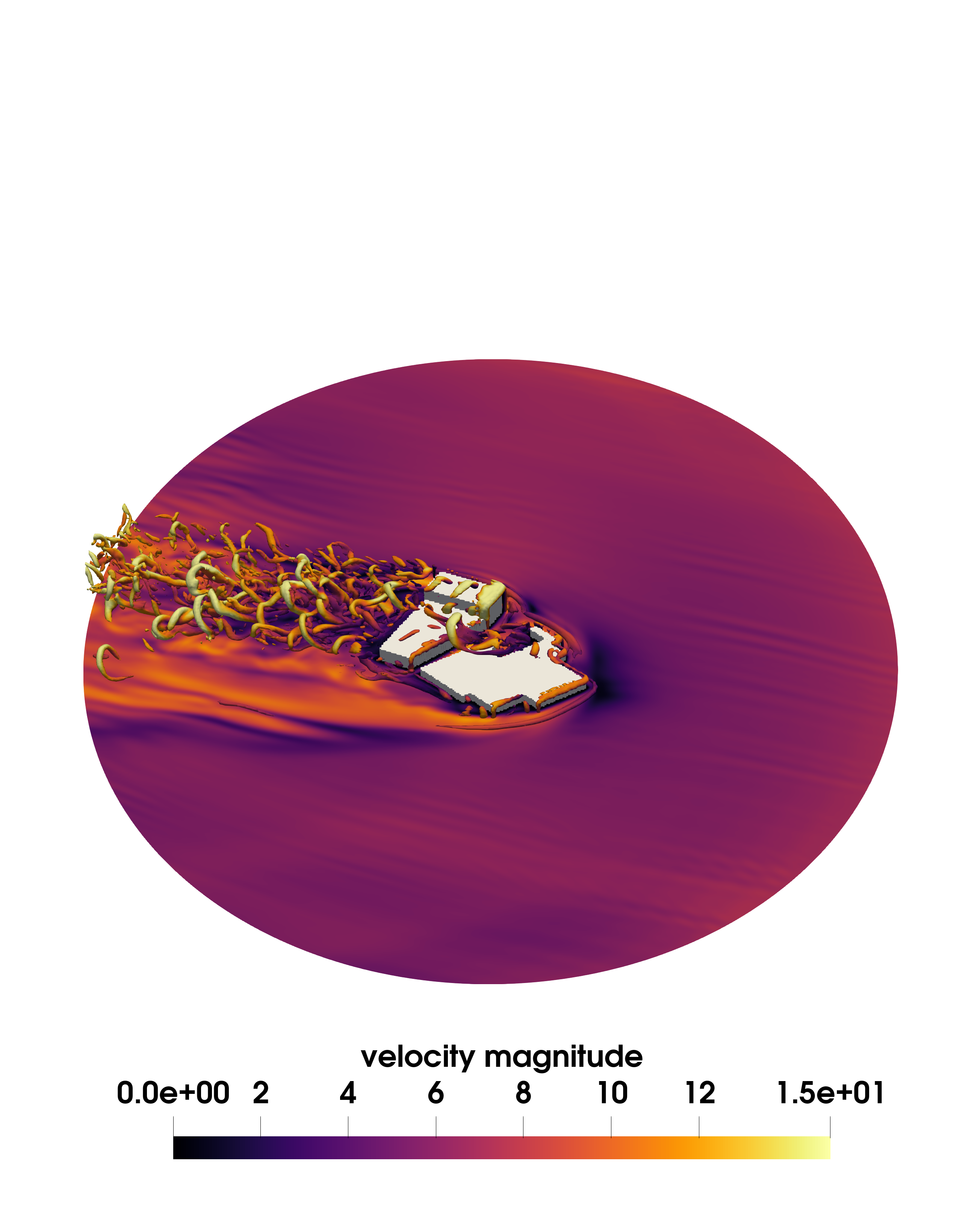}\\
    \subfloat[2024-11-07 22:00:00 CET]{\includegraphics[trim={0 20cm 0 25cm},clip,width=0.32\textwidth]{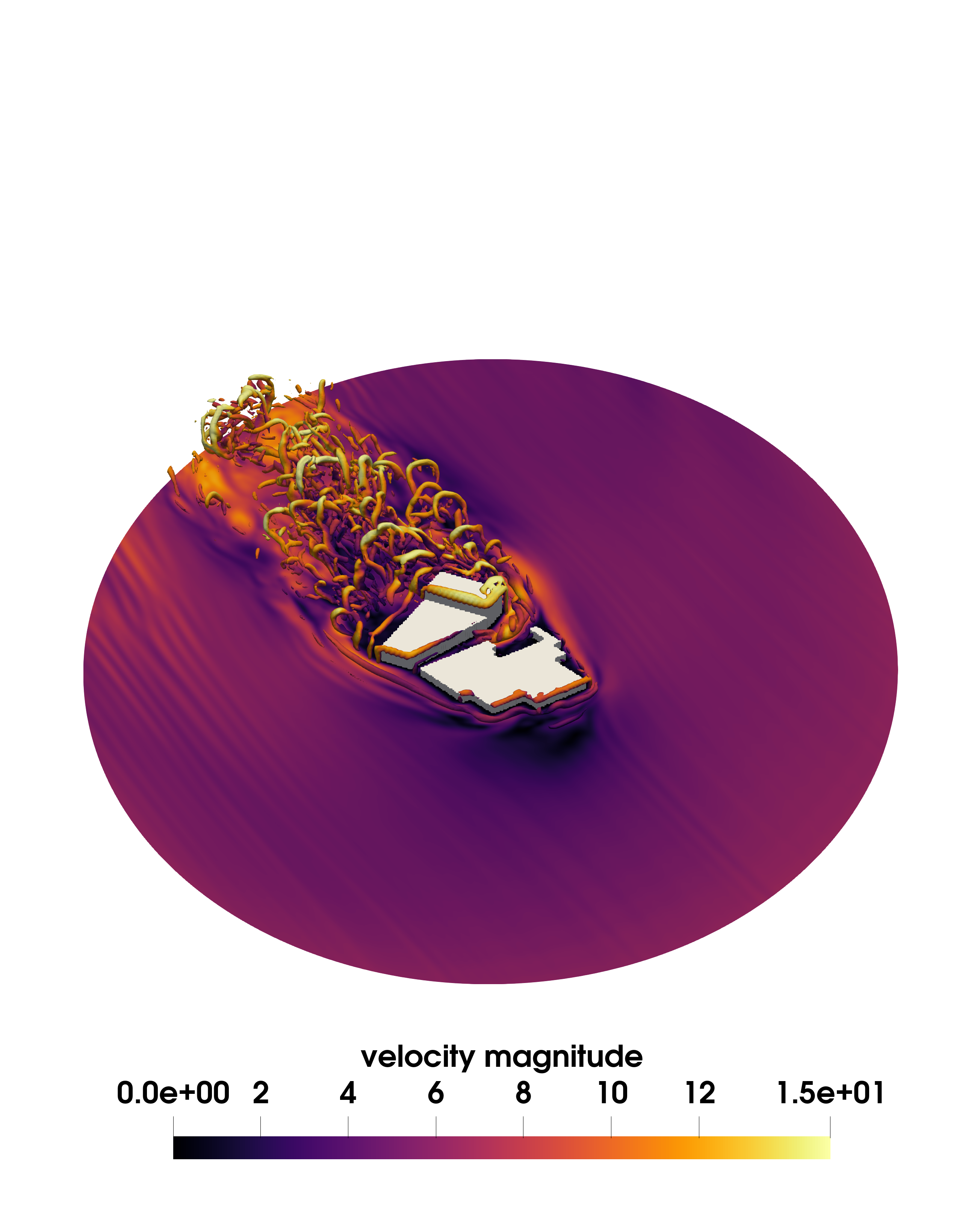}}
    \subfloat[2024-11-08 10:00:00 CET]{\includegraphics[trim={0 20cm 0 25cm},clip,width=0.32\textwidth]{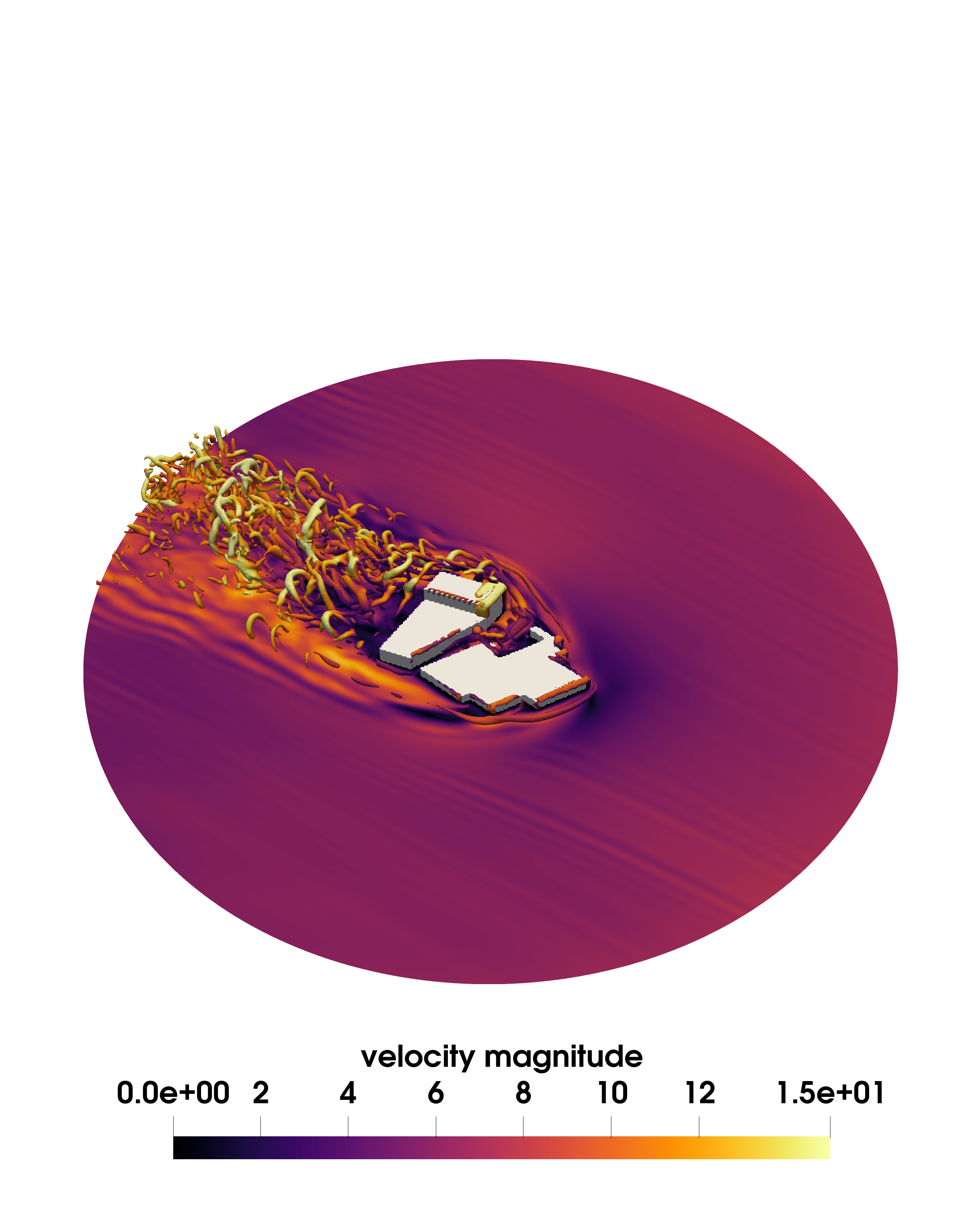}}
    \subfloat[2024-11-08 21:00:00 CET]{\includegraphics[trim={0 20cm 0 25cm},clip,width=0.32\textwidth]{fig/singleSample/magmaLocalVelSingleSample60.0027.png}}\\
    \includegraphics[trim={0 6cm 0 95cm},clip,width=0.5\textwidth]{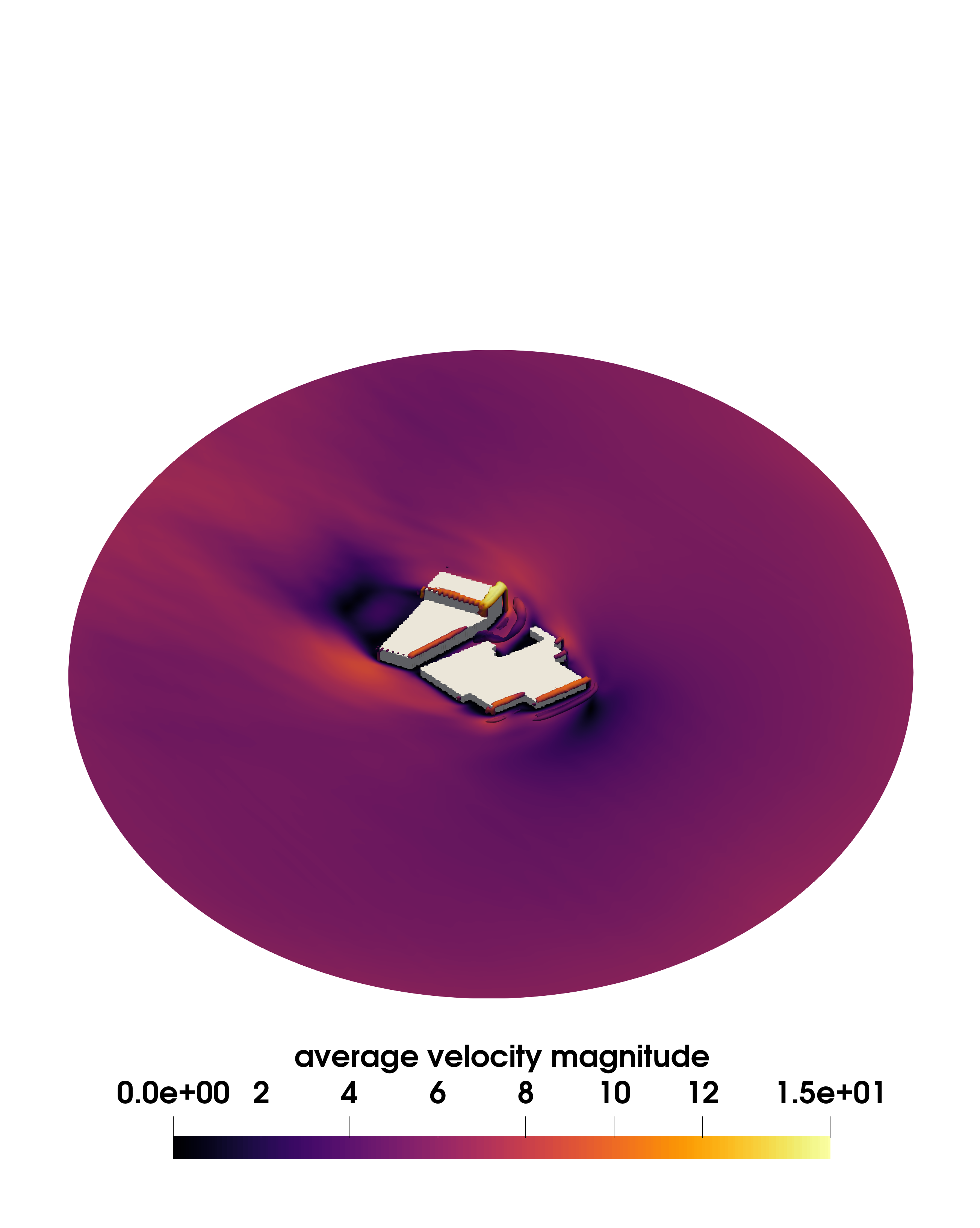}\\
    \subfloat[2024-11-07 22:00:00 CET]{\includegraphics[trim={0 20cm 0 25cm},clip,width=0.32\textwidth]{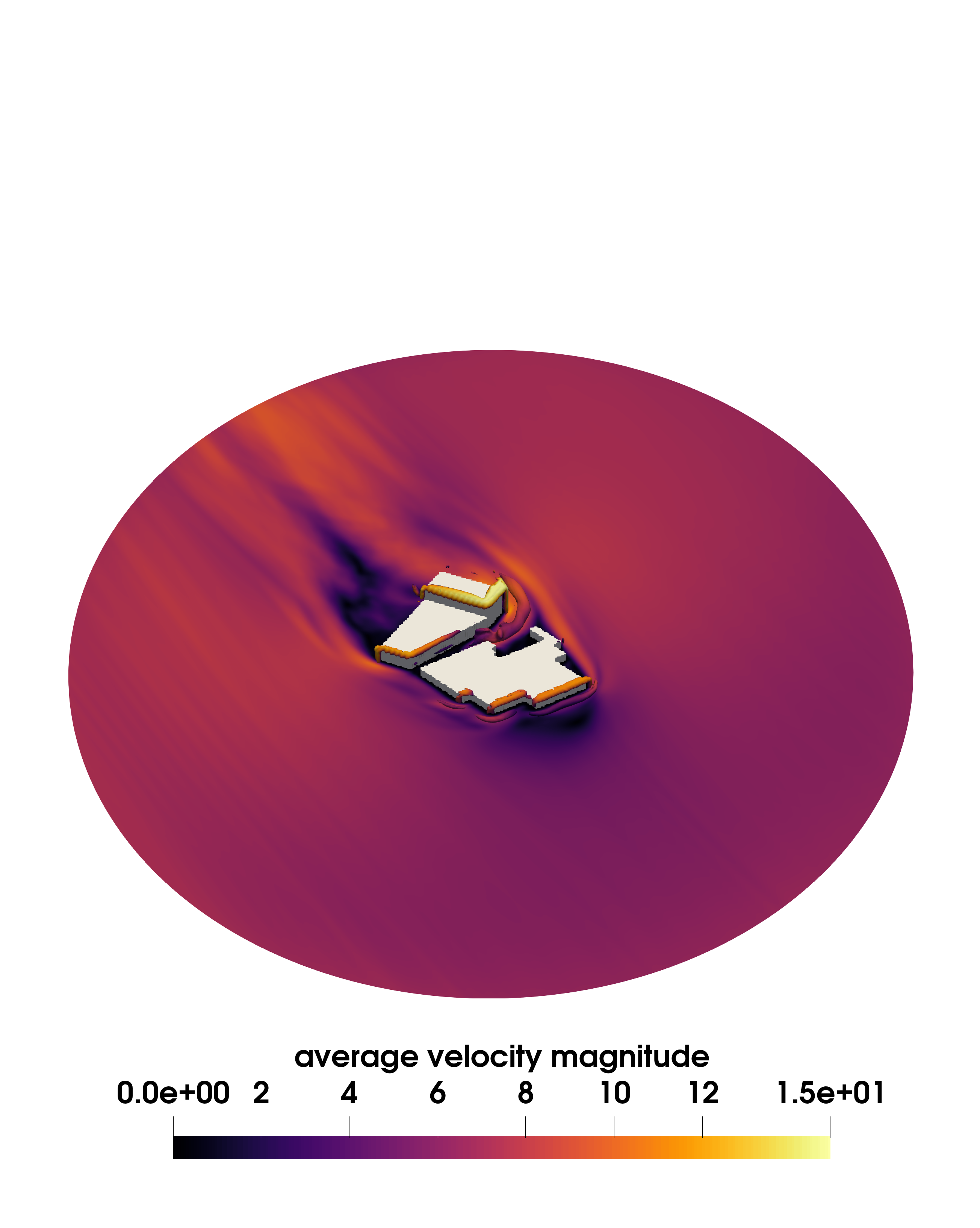}}  
    \subfloat[2024-11-08 10:00:00 CET]{\includegraphics[trim={0 20cm 0 25cm},clip,width=0.32\textwidth]{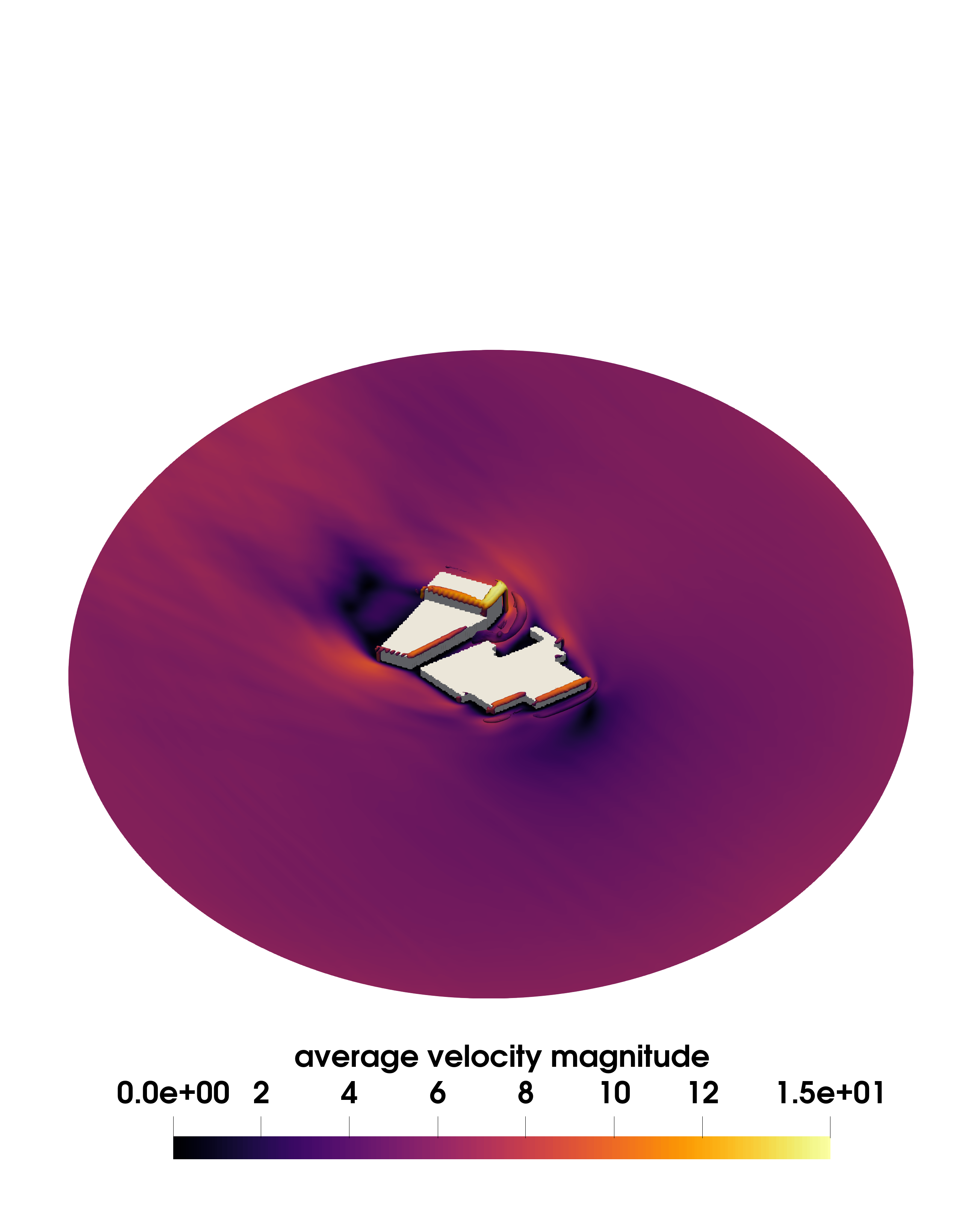}}
    \subfloat[2024-11-08 21:00:00 CET]{\includegraphics[trim={0 20cm 0 25cm},clip,width=0.32\textwidth]{fig/singleSample/magmaAveraVelSingleSample60.0027.png}}
    \caption{
     Simulated (SC LBM) single sample (number 60 out of 121, nearest to expected mean) time-dependent velocity field magnitude at (a) 2024-11-07 22:00:00 CET, (b) 2024-11-08 10:00:00 CET, (c) 2024-11-09 08:00:00, and time-averaged velocity field magnitude at (d) 2024-11-07 22:00:00 CET, (e) 2024-11-08 10:00:00 CET, (f) 2024-11-09 08:00:00, respectively. 
     Isocontours are colored in velocity magnitude and represent the \(Q\)-criterion at \(Q=1\), of the local-in-time results (a,b,c) and the time-averaged time averaged results (d,e,f), respectively.
    }
    \label{fig:single_sample60_Q-crit_local}
\end{figure}

To illustrate the variability across individual samples, we highlight two representative cases from the stochastic collocation ensemble.
Figure~\ref{fig:single_sample10_Q-crit_local} shows results for sample~10 (out of 121), which lies farthest from the expected mean and exhibits a pronounced deviation in terms of instantaneous velocity field structures.
In contrast, Figure~\ref{fig:single_sample60_Q-crit_local} presents sample~60, which lies closest to the expected mean.
For both cases, the time-dependent velocity magnitudes are shown at three selected instants (2024-11-07 22:00:00 CET, 2024-11-08 10:00:00 CET, and 2024-11-09 08:00:00 CET) alongside their corresponding time-averaged fields.
The isocontours of the \(Q\)-criterion (\(Q=1\)) reveal coherent vortical structures, where differences between sample~10 and sample~60 highlight the spread of local flow features induced by input uncertainty.

To further quantify the uncertainty in the simulated flow field, we perform postprocessing over all 121 samples to evaluate the mean and standard deviation of the velocity magnitude.

\begin{figure}[ht!]
    \centering
    \includegraphics[trim={0 0 0 101cm},clip,width=0.5\textwidth]{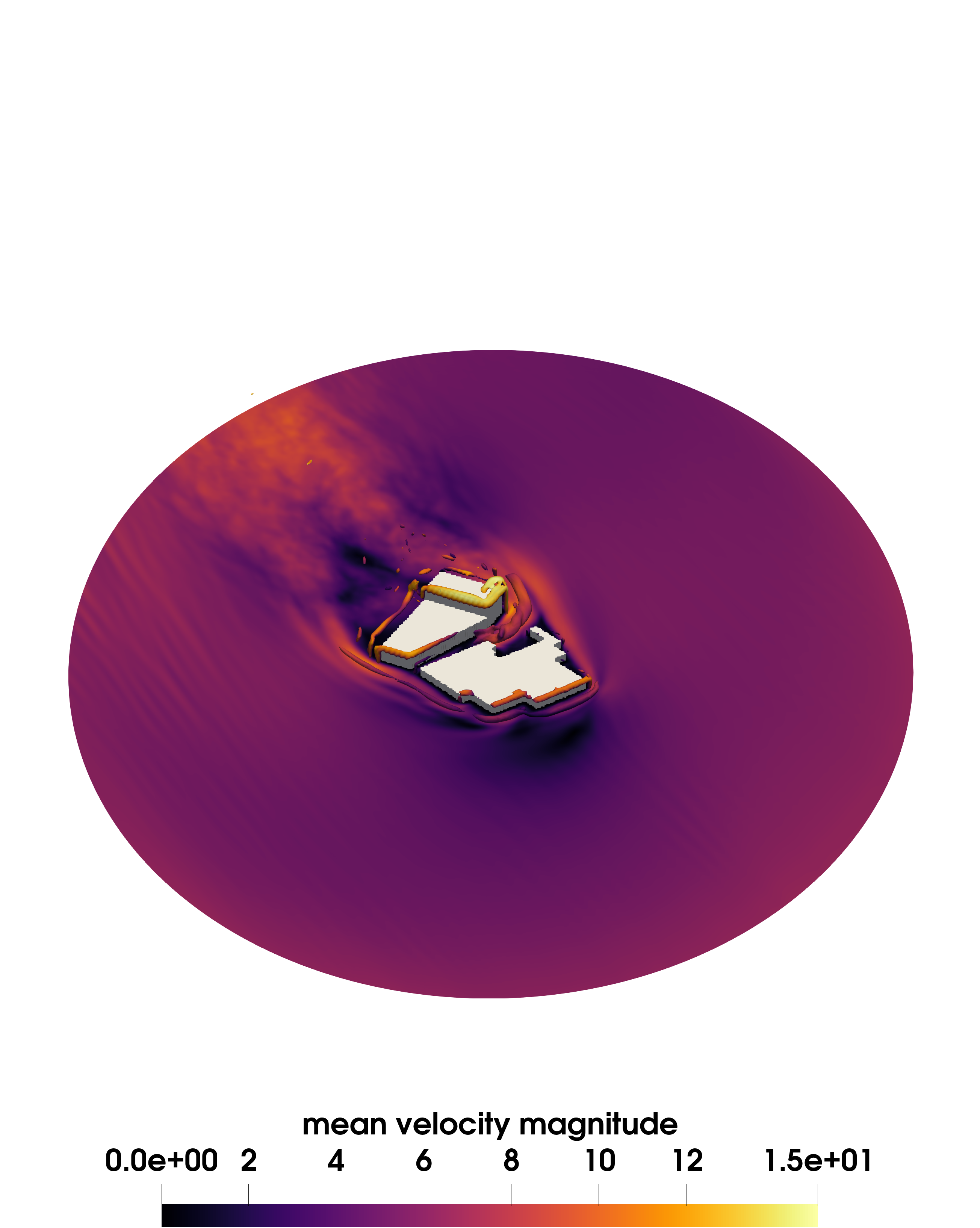}\\
    \subfloat[2024-11-07 22:00:00 CET]{\includegraphics[trim={0 20cm 0 25cm},clip,width=0.32\textwidth]{fig/meanPlots/meanLocalQ.0004.png}}
    \subfloat[2024-11-08 10:00:00 CET]{\includegraphics[trim={0 20cm 0 25cm},clip,width=0.32\textwidth]{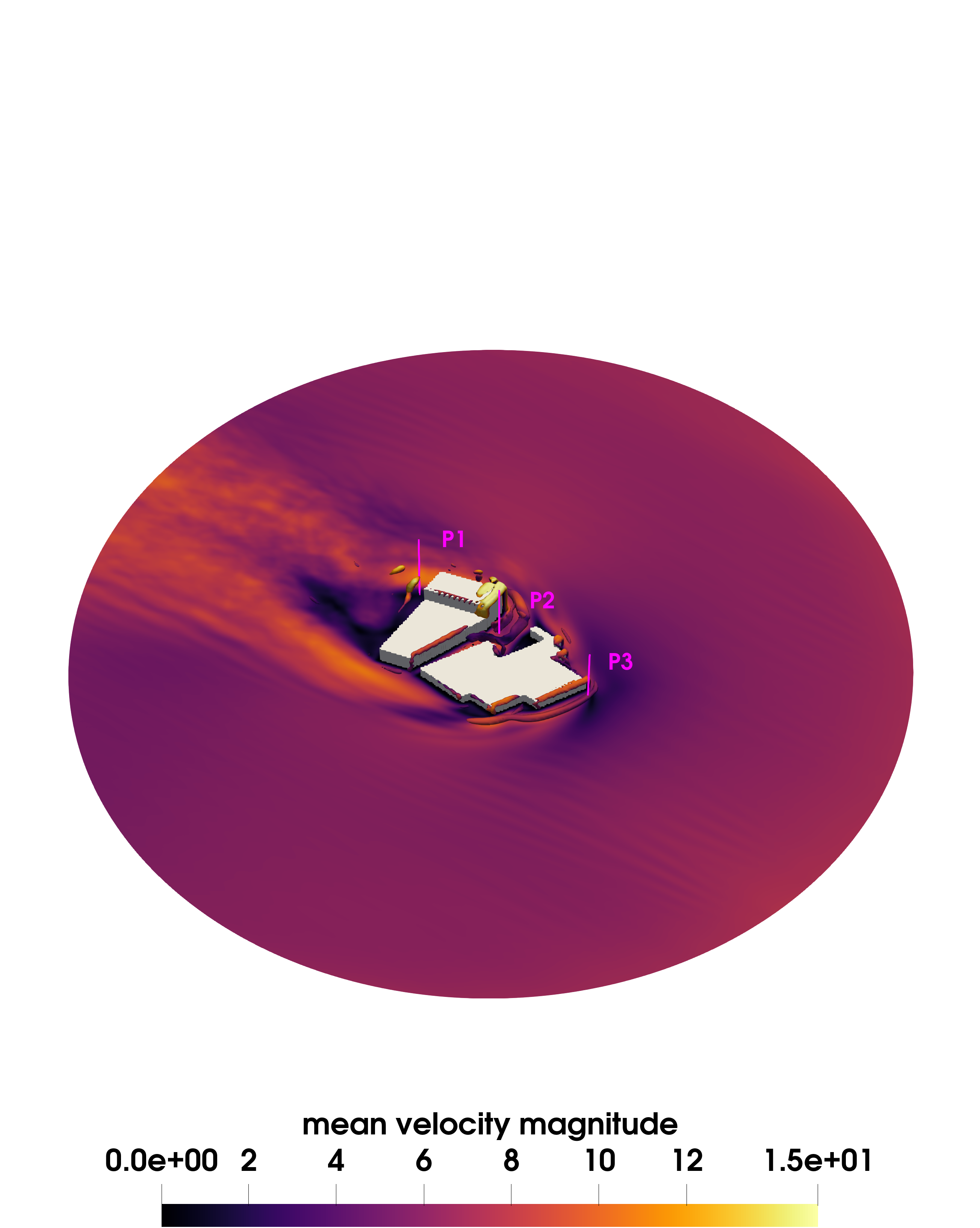}\label{subfig:meanStdLocal_Q_16_probes}}
    \subfloat[2024-11-08 21:00:00 CET]{\includegraphics[trim={0 20cm 0 25cm},clip,width=0.32\textwidth]{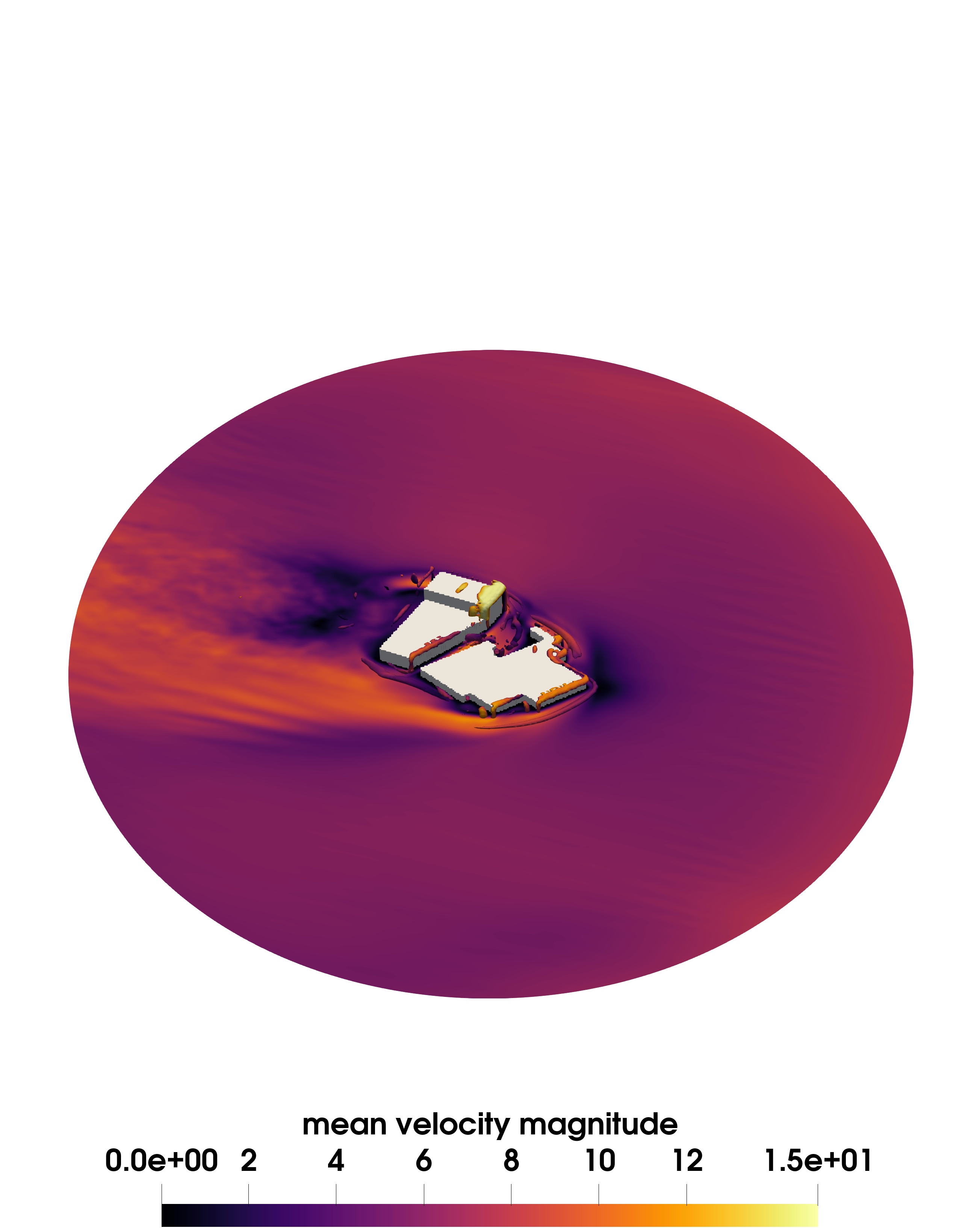}}\\
    \includegraphics[trim={0 0 0 101cm},clip,width=0.5\textwidth]{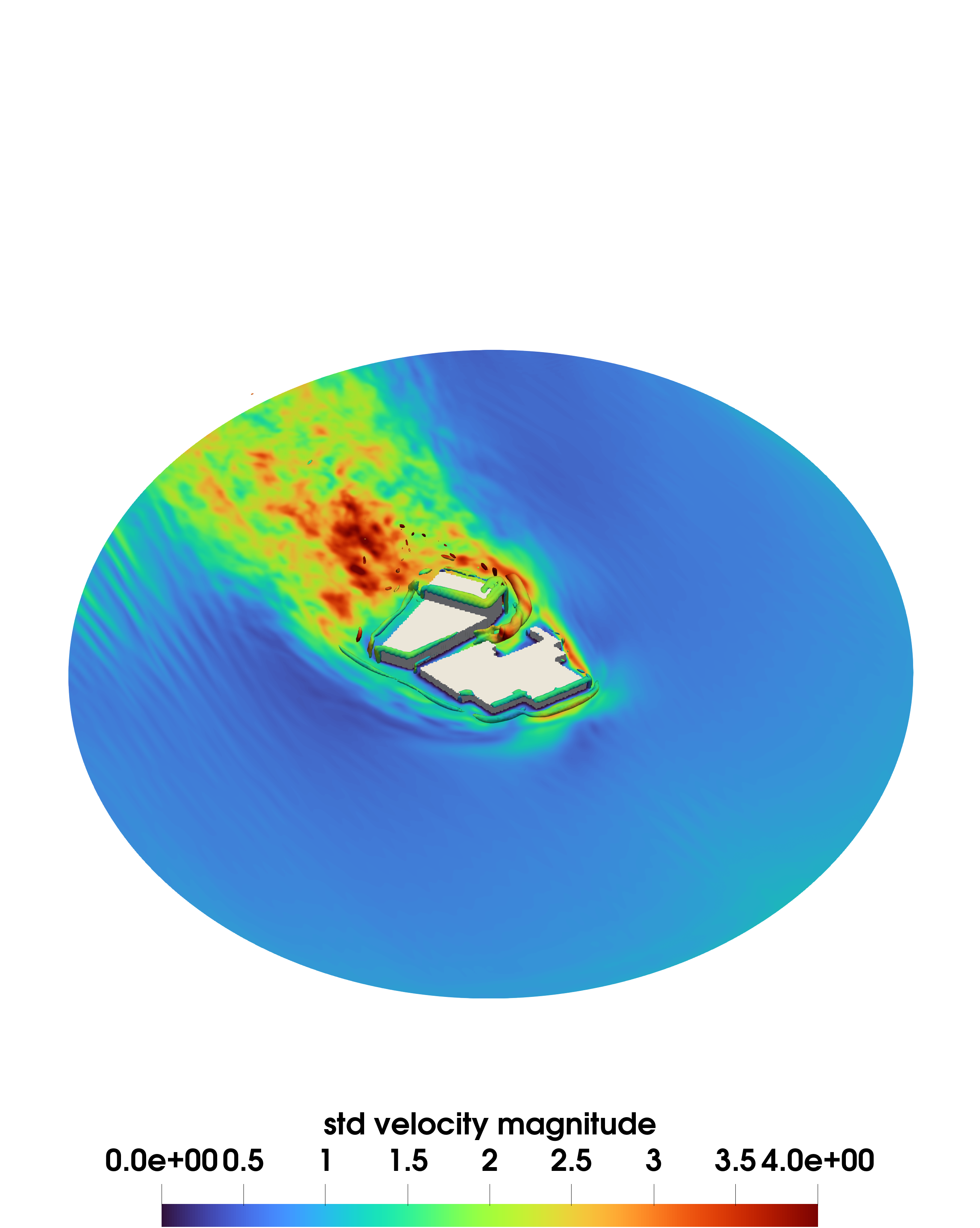}\\
    \subfloat[2024-11-07 22:00:00 CET]{\includegraphics[trim={0 20cm 0 25cm},clip,width=0.32\textwidth]{fig/meanPlots/stdLocalQ.0004.png}}  
    \subfloat[2024-11-08 10:00:00 CET]{\includegraphics[trim={0 20cm 0 25cm},clip,width=0.32\textwidth]{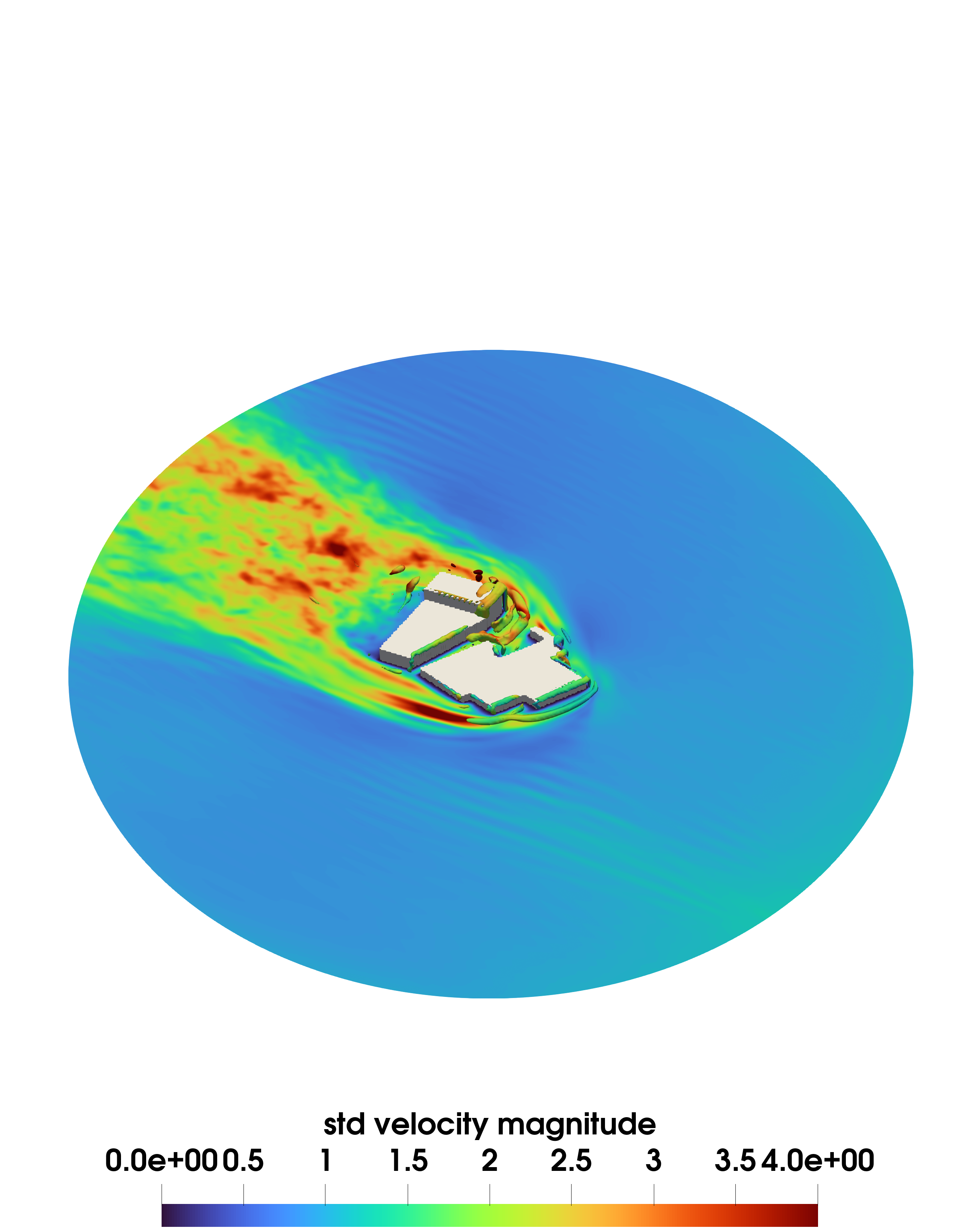}}
    \subfloat[2024-11-08 21:00:00 CET]{\includegraphics[trim={0 20cm 0 25cm},clip,width=0.32\textwidth]{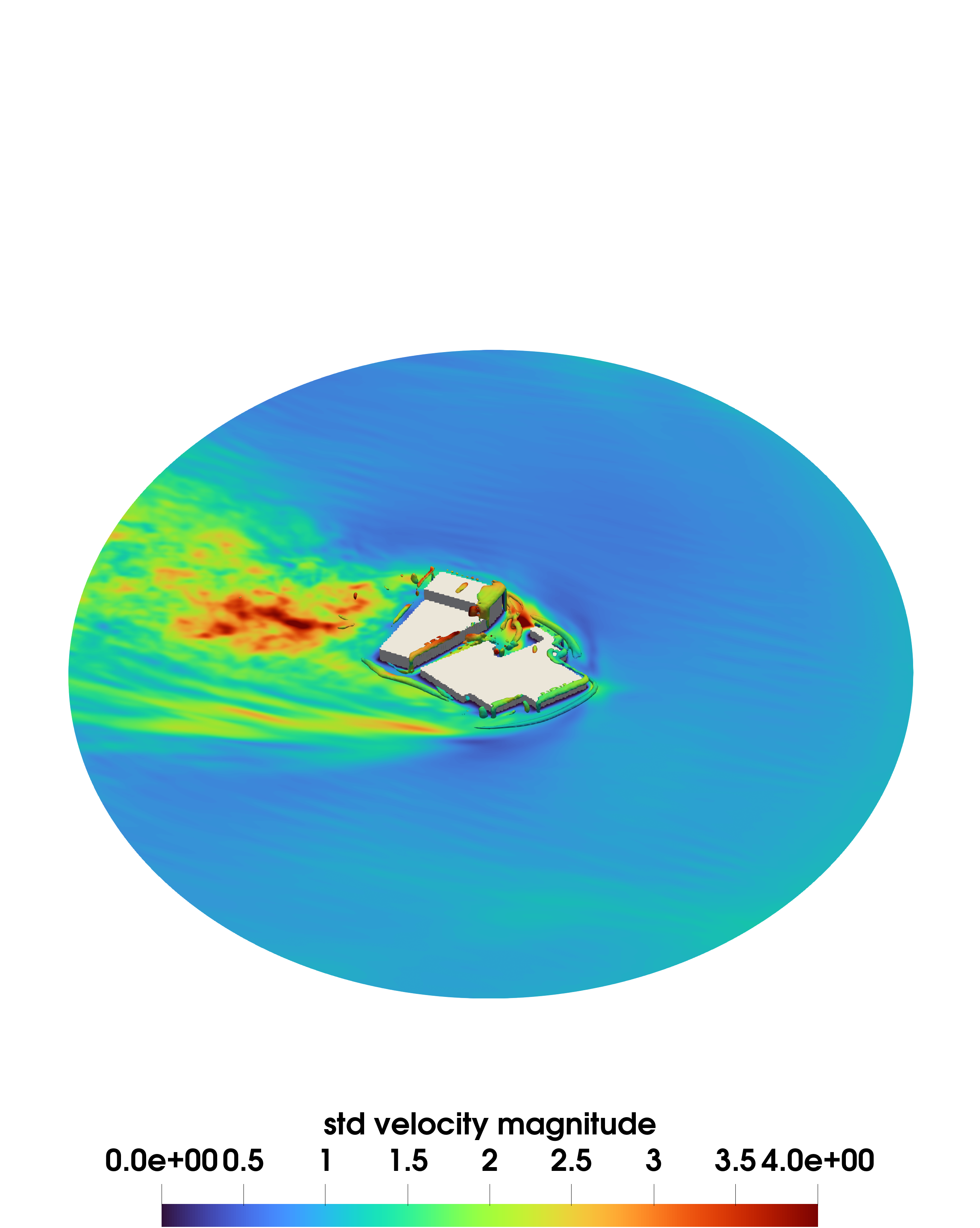}}
    \caption{
     Simulated (SC LBM) mean of the time-dependent velocity field magnitude at (a) 2024-11-07 22:00:00 CET, (b) 2024-11-08 10:00:00 CET, (c) 2024-11-09 08:00:00, and std of the time-dependent velocity field magnitude at (d) 2024-11-07 22:00:00 CET, (e) 2024-11-08 10:00:00 CET, (f) 2024-11-09 08:00:00, respectively, both computed from 121 samples. 
     Isocontours are computed from mean velocity magnitudes and represent the \(Q\)-criterion at \(Q=1\). 
     The isocontours are respectively colored in mean (a,b,c) and std (d,e,f) of the time-dependent velocity magnitude. 
     The probe probe location lines P1, P2, P3 are shown in pink (b), further specified in Table~\ref{tab:probes}, and evaluated in Figure~\ref{fig:urban_probes_profiles}.
    }
    \label{fig:meanStdLocal_Q}
\end{figure}

Figure~\ref{fig:meanStdLocal_Q} shows the statistics of the time-dependent velocity magnitude at three representative instants (2024-11-07 22:00:00~CET, 2024-11-08 10:00:00~CET, and 2024-11-09 08:00:00~CET). The mean fields (a--c) capture the predominant flow structures around the buildings, whereas the corresponding standard deviation fields (d--f) highlight regions of elevated variability, particularly in the wakes and shear layers.

\begin{figure}[ht!]
    \centering
    \includegraphics[trim={0 0 0 101cm},clip,width=0.5\textwidth]{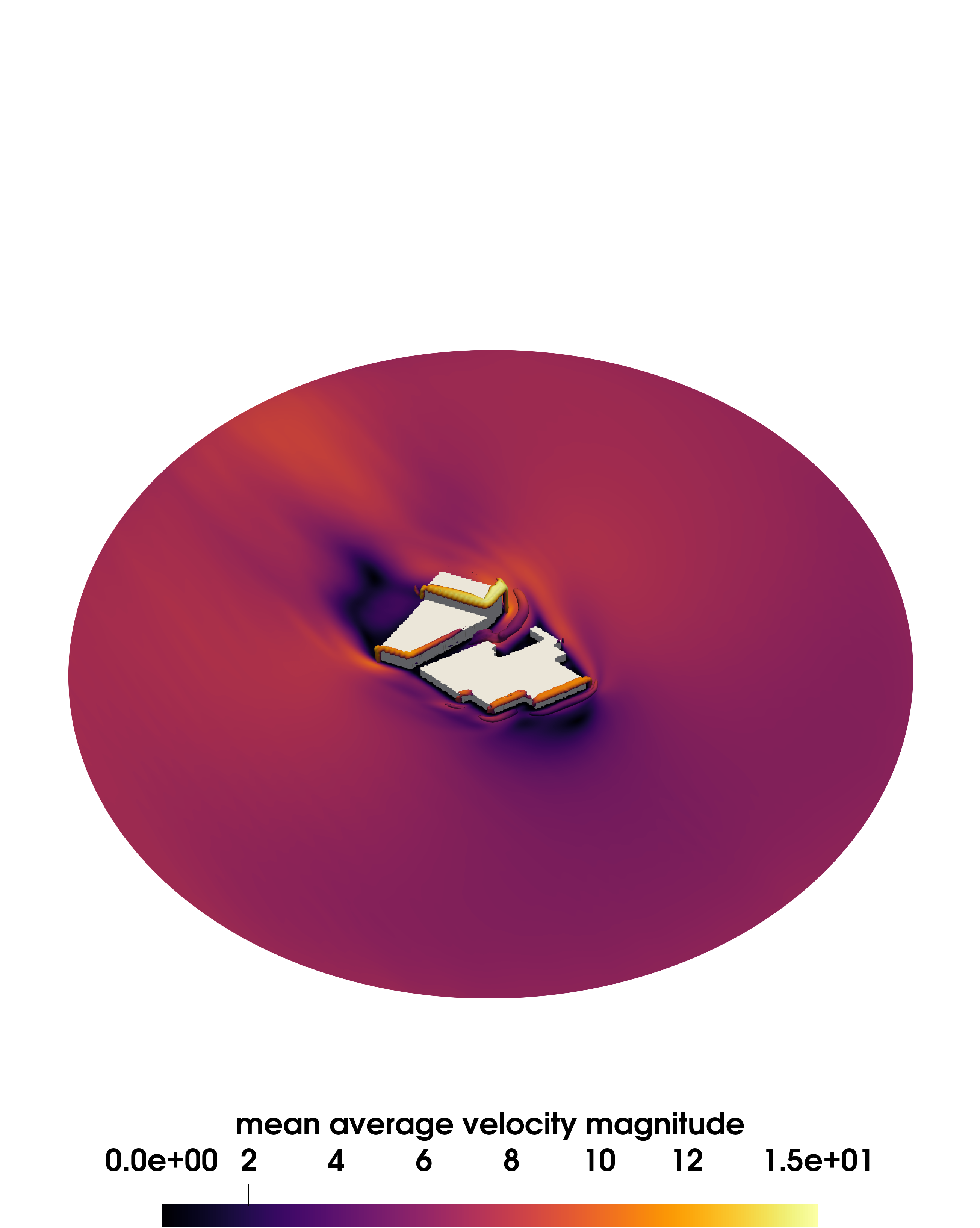}\\
    \subfloat[2024-11-07 22:00:00 CET]{\includegraphics[trim={0 20cm 0 25cm},clip,width=0.32\textwidth]{fig/meanPlots/meanAverageQ.0004.png}}
    \subfloat[2024-11-08 10:00:00 CET]{\includegraphics[trim={0 20cm 0 25cm},clip,width=0.32\textwidth]{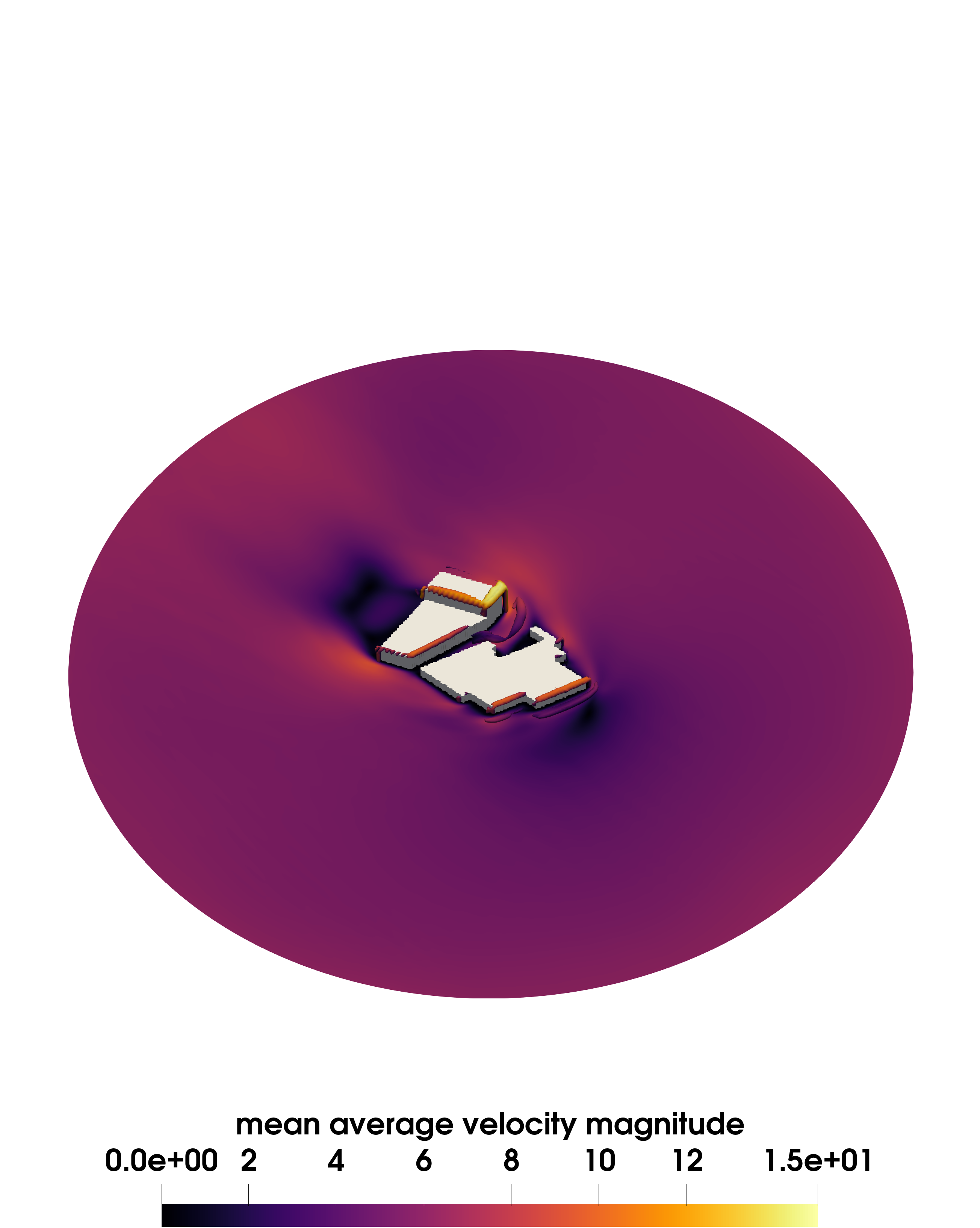}}
    \subfloat[2024-11-08 21:00:00 CET]{\includegraphics[trim={0 20cm 0 25cm},clip,width=0.32\textwidth]{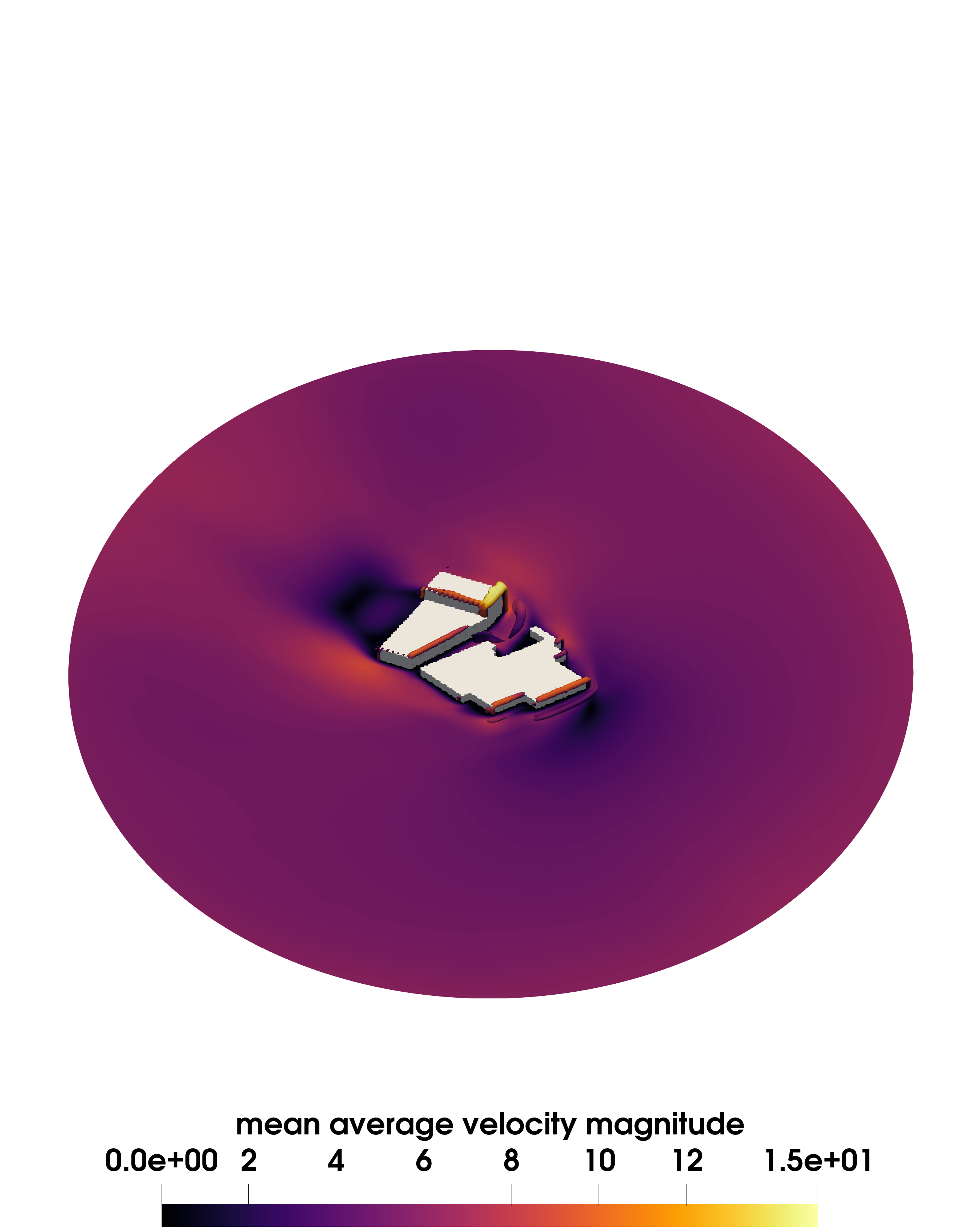}}\\
    \includegraphics[trim={0 0 0 101cm},clip,width=0.5\textwidth]{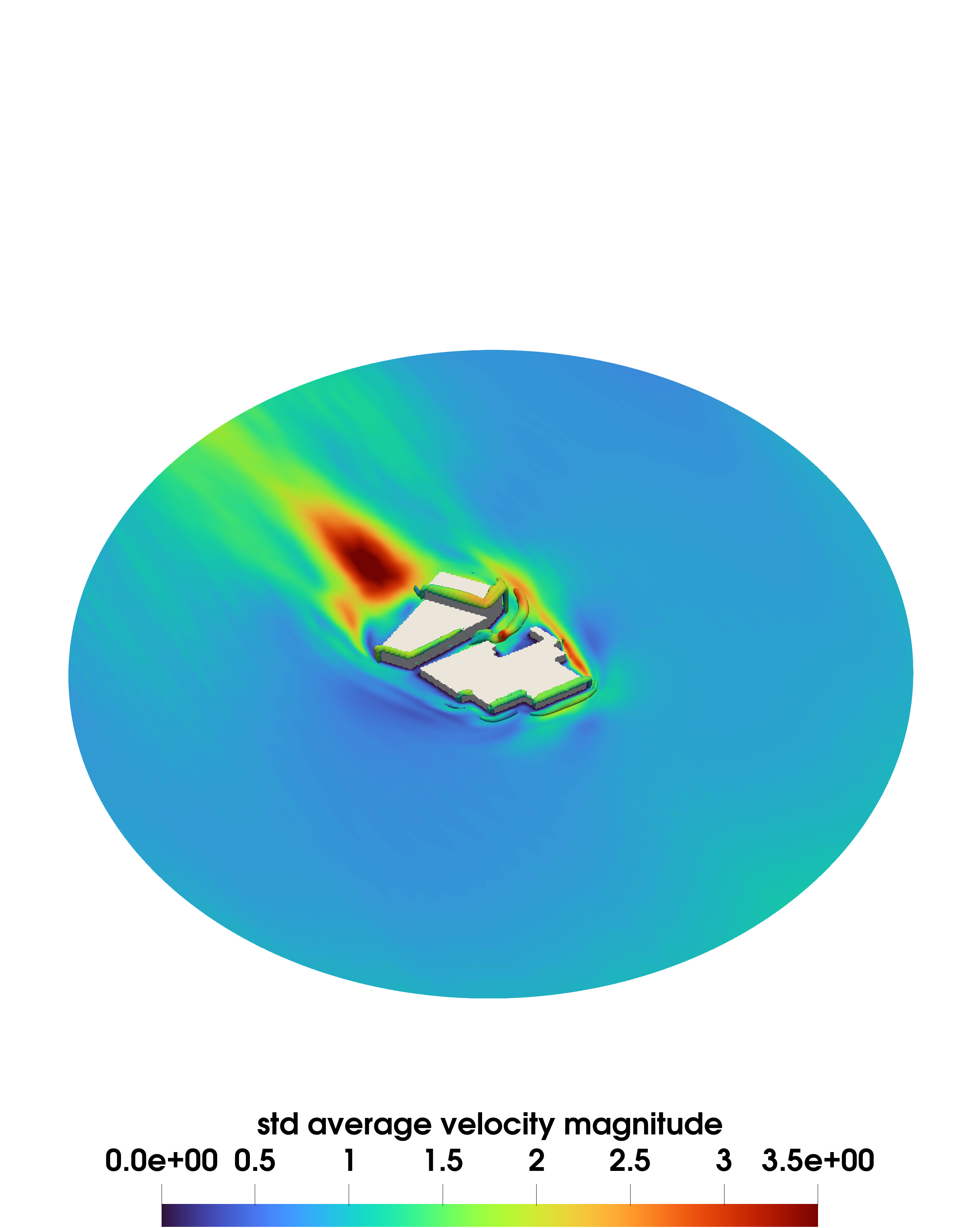}\\
    \subfloat[2024-11-07 22:00:00 CET]{\includegraphics[trim={0 20cm 0 25cm},clip,width=0.32\textwidth]{fig/meanPlots/stdAverageQ.0004.png}}  
    \subfloat[2024-11-08 10:00:00 CET]{\includegraphics[trim={0 20cm 0 25cm},clip,width=0.32\textwidth]{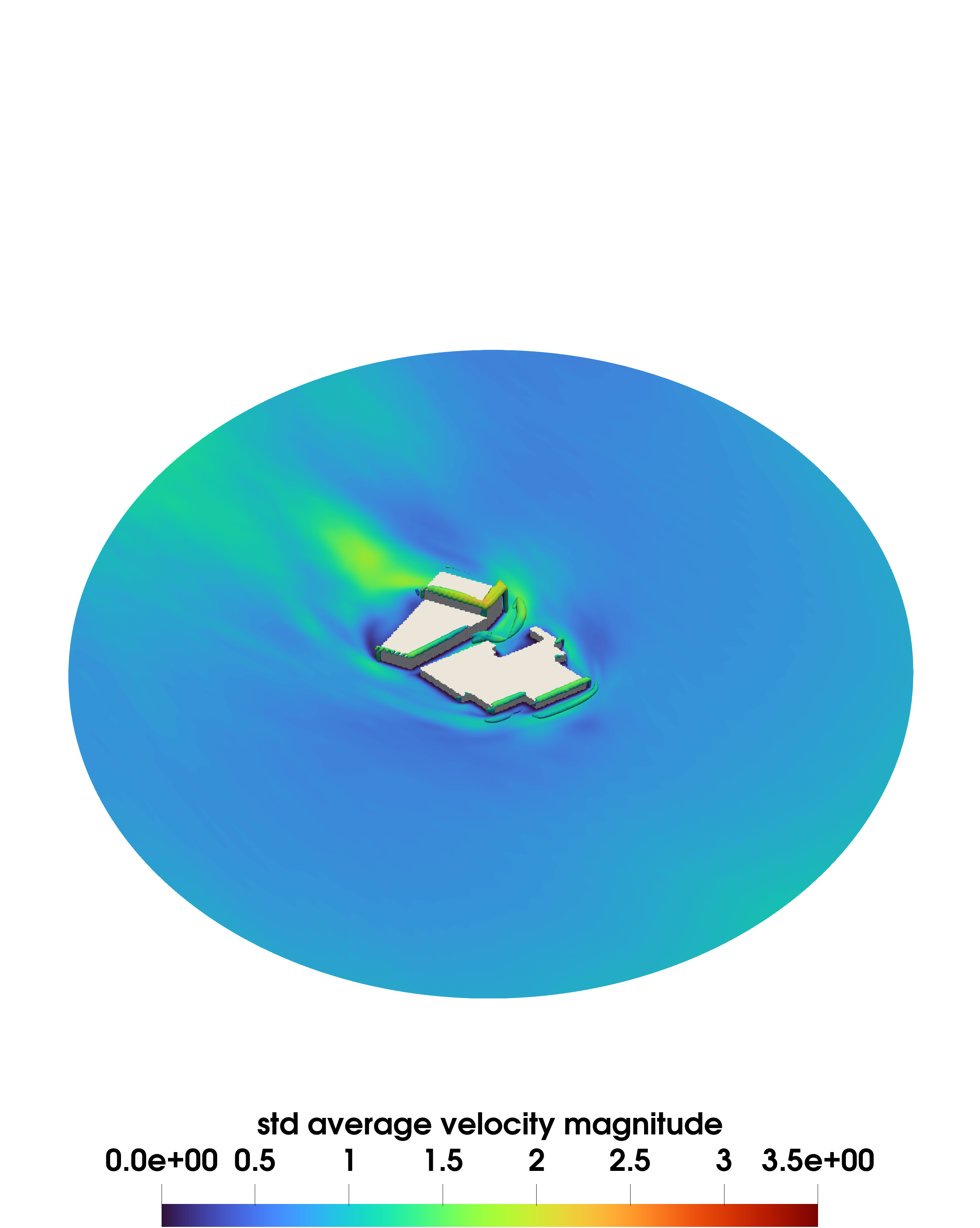}}
    \subfloat[2024-11-08 21:00:00 CET]{\includegraphics[trim={0 20cm 0 25cm},clip,width=0.32\textwidth]{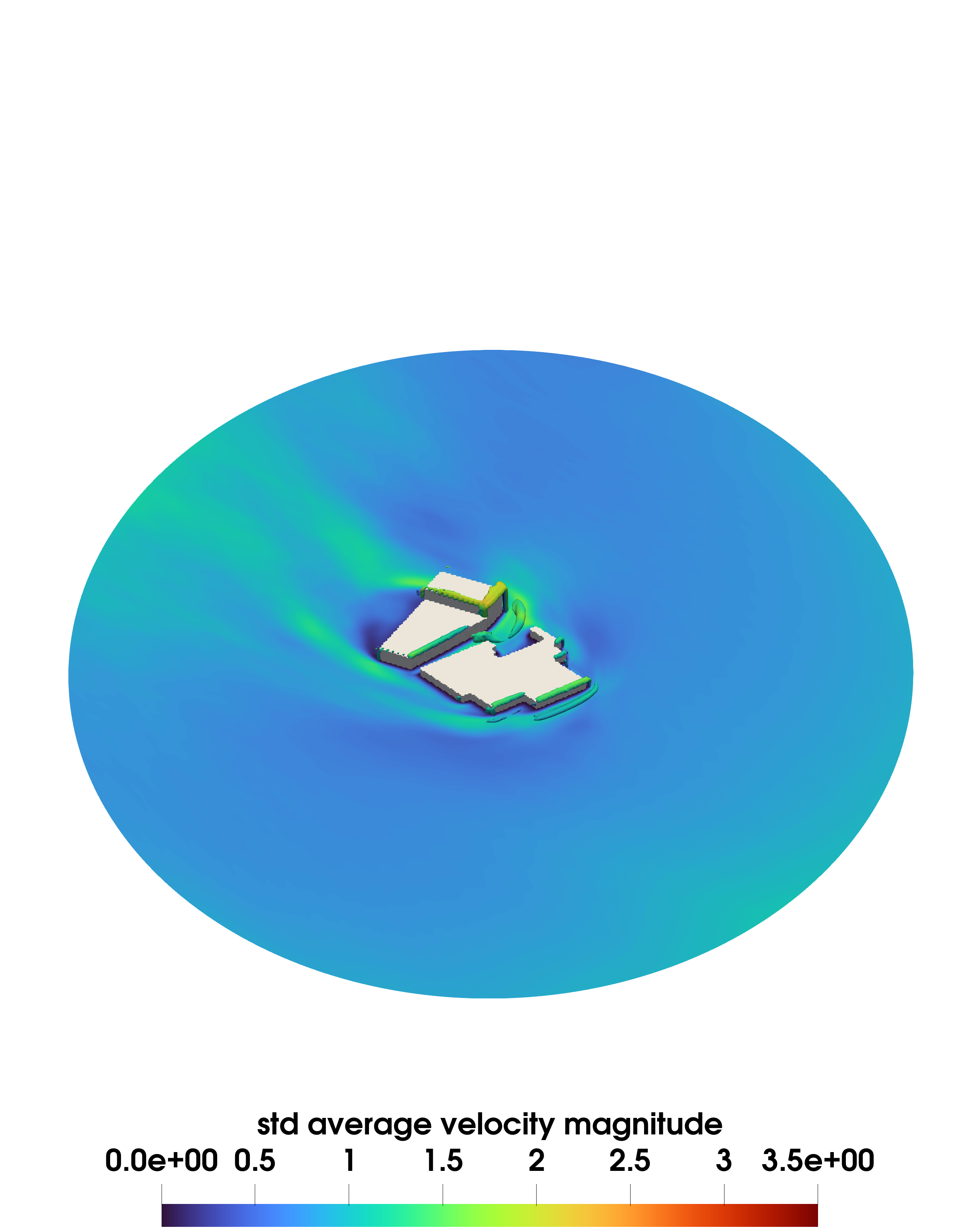}}
    \caption{
     Simulated (SC LBM) mean of the time-averaged velocity field magnitude at (a) 2024-11-07 22:00:00 CET, (b) 2024-11-08 10:00:00 CET, (c) 2024-11-09 08:00:00, and std of the time-averaged velocity field magnitude at (d) 2024-11-07 22:00:00 CET, (e) 2024-11-08 10:00:00 CET, (f) 2024-11-09 08:00:00, respectively, both computed from 121 samples. 
     Isocontours are computed from mean velocity magnitudes and represent the \(Q\)-criterion at \(Q=1\). 
     The isocontours are respectively colored in mean (a,b,c) and std (d,e,f) of the time-averaged velocity magnitude. 
    }
    \label{fig:meanStdAverage_Q}
\end{figure}

Complementing this, Figure~\ref{fig:meanStdAverage_Q} presents the mean and standard deviation of the time-averaged velocity magnitude. Here, the mean field emphasizes the dominant flow patterns, while the standard deviation field identifies persistent regions of high variability. Together, these statistical characterizations provide a comprehensive picture of the flow, offering insight into the reliability of the simulations and pinpointing regions most sensitive to input uncertainty.

To further investigate the vertical wind profile and its uncertainty at selected positions in the urban domain, 
we place three virtual probes at the coordinates listed in Table~\ref{tab:probes} (see also pink points in Figure~\ref{fig:meanStdLocal_Q}), representing distinct flow regions.
\begin{table}[ht!]
\centering
\footnotesize
\caption{Probe locations and associated flow regions in the urban domain (see also black line markers in Figure~\ref{subfig:meanStdLocal_Q_16_probes}). The computed vertical velocity magnitude profiles at these probe locations are shown in Figure~\ref{fig:urban_probes_profiles}.}
\label{tab:probes}
\begin{tabular}{cr}
\hline
\makecell[c]{\textbf{Probe specifier}} & \makecell[c]{\textbf{Coordinates} $(x, y,z)$} \\
\hline
P1 & $(30, 87, z), ~ z\in [0,40]$ \\
P2 & $(78, 60, z), ~ z\in [0,40]$ \\
P3 & $(130, 15, z), ~ z\in [0,40]$ \\
\hline
\end{tabular}
\end{table} 

\begin{figure}[ht!]
  \centering
  \includegraphics[width=\textwidth]{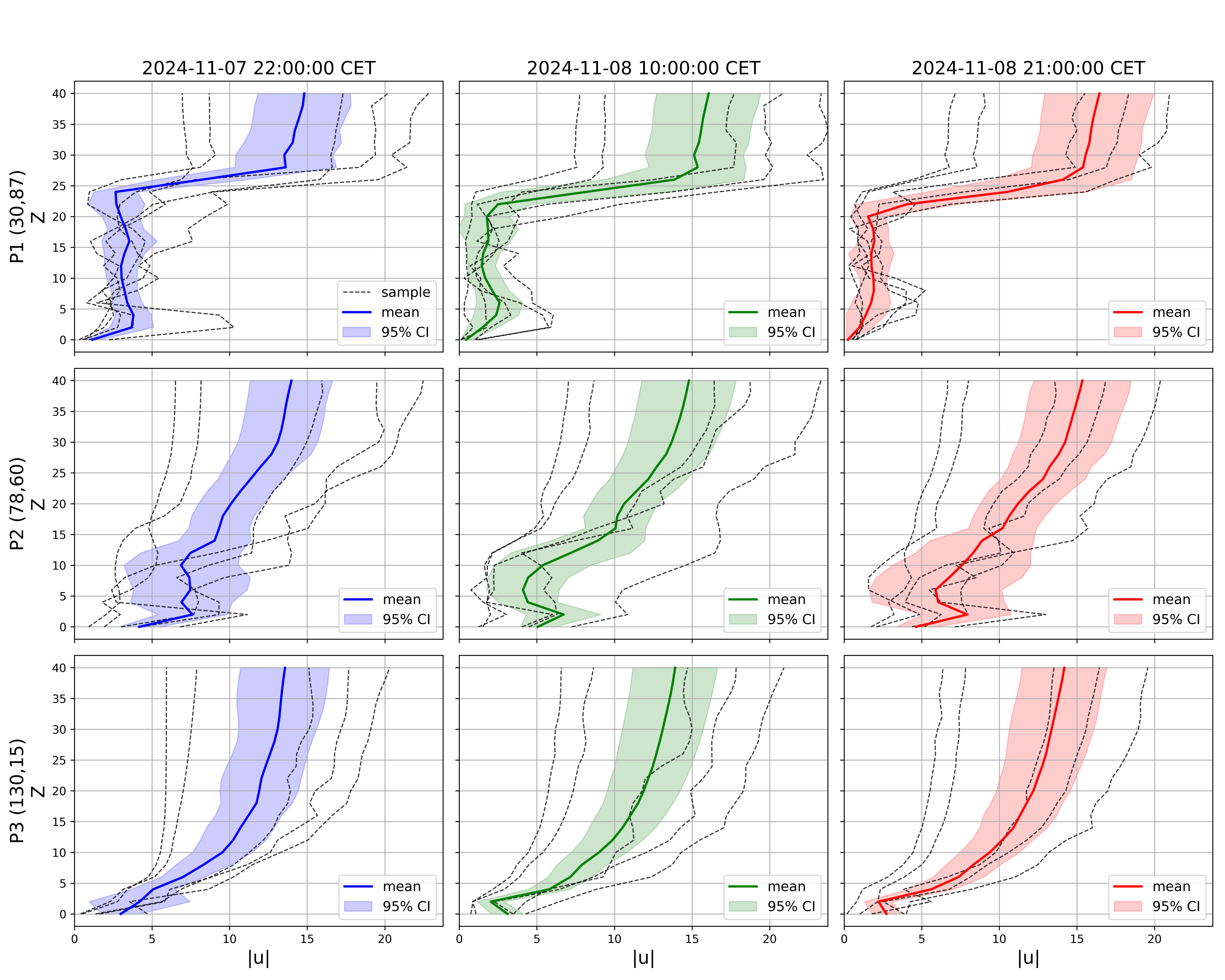}
  \caption{
    Time-local vertical velocity magnitude profiles at three probe positions: 
    P1\,\((30, 87)\) at the outer corner of the left building (shear layer), 
    P2\,\((78, 60)\) inside the channel between the buildings (channel), and 
    P3\,\((130, 15)\) at the outer corner of the right building (shear layer) 
    (see Table~\ref{tab:probes} and Figure~\ref{subfig:meanStdLocal_Q_16_probes}).
    Each subplot corresponds to a specific time: 
    2024-11-07 22:00:00 CET, 
    2024-11-08 10:00:00 CET, and 
    2024-11-08 21:00:00 CET, respectively.
    Solid lines show the mean velocity profile, shaded regions indicate the 95\% confidence intervals, and dashed lines represent five exemplary individual samples (out of 121) from the stochastic collocation LBM simulation.
    This figure visualizes one selected time step per subplot.  
  }
  \label{fig:urban_probes_profiles}
\end{figure}

Figure~\ref{fig:urban_probes_profiles} shows the vertical distribution of the velocity magnitude at each probe, including the mean, 95\% CI, and individual sample realizations obtained from the SC LBM simulation. 
The width of the confidence bands varies significantly between locations, indicating different levels of flow variability depending on the local building-induced effects and turbulent mixing intensity.

\begin{figure}[ht!]
  \centering
  \subfloat[mean average at 2024-11-08 21:00:00 CET]{
    \includegraphics[width=.45\linewidth]{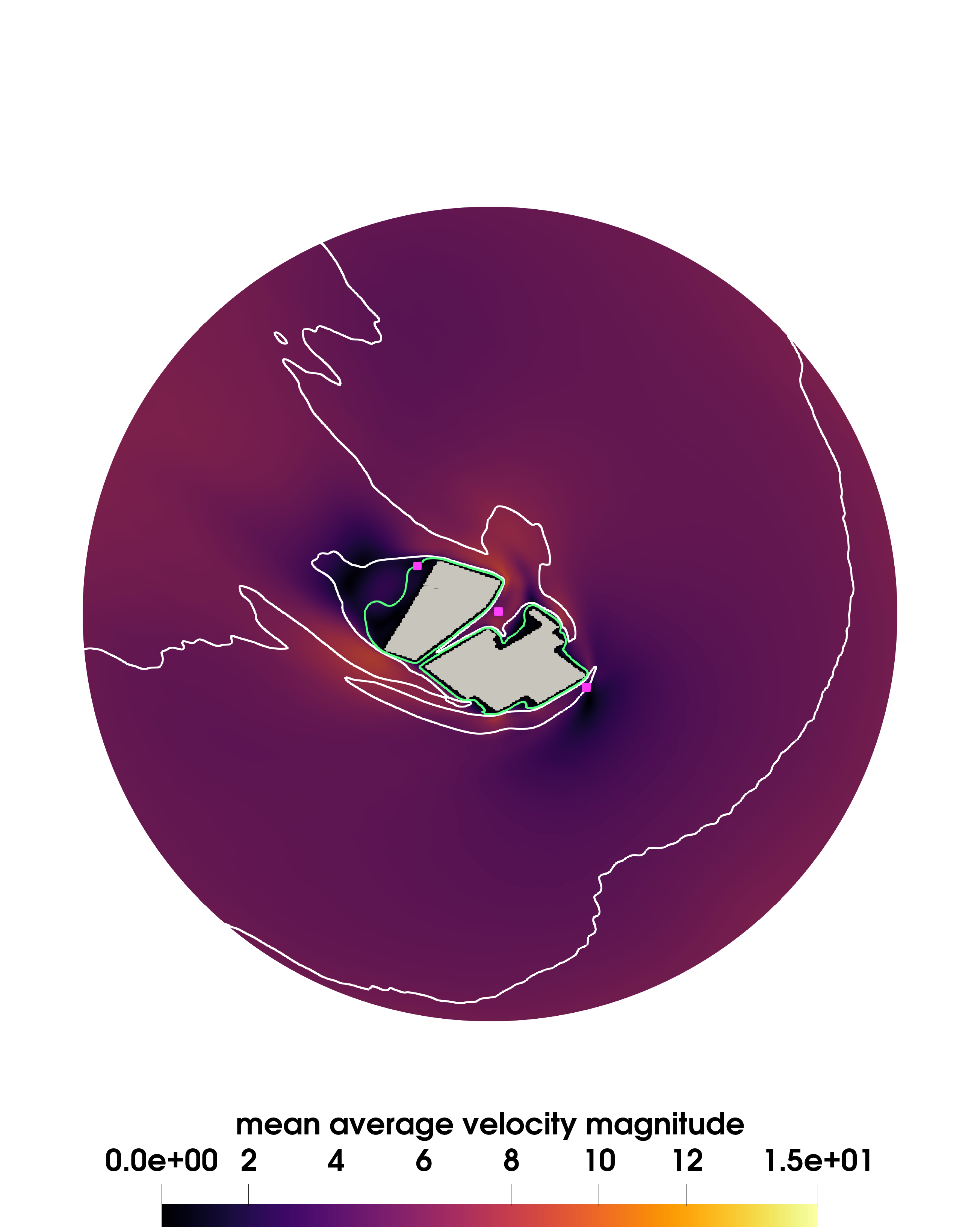}\label{subfig:urban_mean_std_contour_average}}
  \subfloat[mean at 2024-11-08 21:00:00 CET]{
      \includegraphics[width=.45\linewidth]{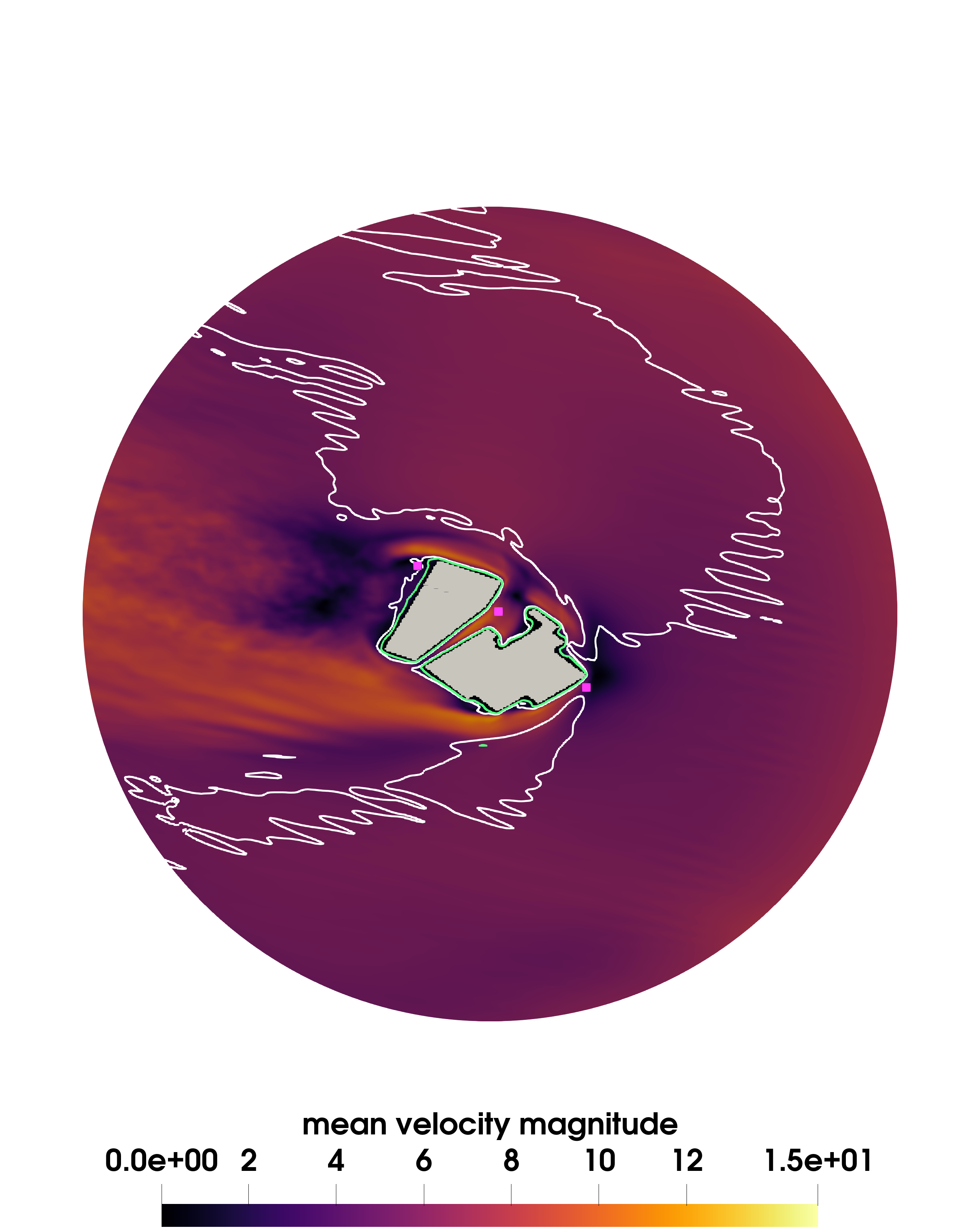}\label{subfig:urban_mean_std_contour_local}}
  \caption{Horizontal slice of the urban domain at a height of $2\,\mathrm{m}$ above ground, showing the mean velocity magnitude field (background colormap), for (a) time-averaged and (b) time-dependent results, with overlaid iso-contours of the velocity standard deviation at $\sigma_u \approx 0.8$ ($50\%$ in white) and $\sigma_u \approx 0.4$ ($25\%$ in green). The standard deviation in this slice ranges from $0$ to $\approx 1.63$, further decreases inside the isocounters. The probe locations P1--P3 are visualized as pink dots and further specified in Table~\ref{tab:probes}.
  }
  \label{fig:urban_mean_std_contour}
\end{figure}

Figure~\ref{fig:urban_mean_std_contour} shows a horizontal slice of the urban domain at a height of $2\,\mathrm{m}$ above ground level.
The background colormap represents the mean velocity magnitude (time-averaged field in Fig.~\ref{subfig:urban_mean_std_contour_average} and time-dependent in Fig.~\ref{subfig:urban_mean_std_contour_local} ), while two iso-contours highlight regions of reduced variability: the white contour corresponds to $\sigma_u \approx 0.8 $ and the green contour to $\sigma_u \approx 0.4$, which refer to $50\%$ and $25\%$ of the absolute standard deviation, respectively.
In this slice, the standard deviation ranges from $0$ to $1.63$, decreasing outside the white contour and further increasing inside the green contour. 
The other regions of enhanced fluctuations are primarily associated with wake zones, shear layers, and open channels between buildings.  
The probe locations P1--P3 are also indicated as pink dots.
Notably, P1 is located in a flow region with an expected mean velocity magnitude smaller than 50\% of the maximum and a standard deviation that is below 25\%.
Thus, based on our simulation results, we are also able to isolate flow regions that have critical features, e.g.\ wind velocity magnitude, up to a specific certainty. 

\section{Conclusion}
\label{sec:conclusion}

We presented an uncertain data assimilation workflow based on OpenLB-UQ~\cite{zhong2025openlbuq}, which directly injects measurement uncertainty into boundary data and propagates it through an SC LBM simulation of the urban wind flow around an isolated real building geometry in the city of Reutlingen (Germany, \(48.4914^\circ\mathrm{N}, 9.2043^\circ\mathrm{E}\)).  
The resulting stochastic boundary condition is mapped to a dense quadrature grid in the gPC space, and OpenLB-UQ is used to compute an ensemble of LBM solutions at the collocation nodes.
From these, we recover spatio-temporal statistics of wind speed and velocity magnitude throughout the urban domain, including diagnostics tailored to urban applications. 
In summary, by modeling inflow uncertainty as a relative perturbation and propagating it through a non-intrusive SC LBM pipeline, we achieved the following:
\begin{itemize}\setlength{\itemsep}{0.0em}
  \item An efficiently scalable first application case of a combined UQ LES framework for urban wind flow simulations in real geometries.
  \item Space-time-dependent estimates of measurement data-assimilated mean and standard deviation of the velocity field in a real urban environment;
  \item Space-time-dependent resolved uncertainty diagnostics highlighting flow-sensitive zones (e.g., wakes, shear layers); 
  \item Confidence intervals at selected monitoring probes.
\end{itemize}

Conclusively, our workflow respects the statistical nature of measurement-driven boundary data and provides interpretable uncertainty maps for further use in downstream applications such as urban design and planning.
Promising future research directions should extend our methodology to handle multi-modal distributions, incorporate machine learning surrogate models (e.g., variational autoencoder), and enable real-time integration with sensor data streams for live urban wind assessment.





\section*{Data availability}
All computations have been done with OpenLB-UQ (also unreleased) on commit hash \href{https://gitlab.com/openlb/olb/-/commit/3953b0ece2caca78b10c0c8075119f0f23ce8fb1}{3953b0ec}.
The OpenLB-UQ framework used in this paper is part of the OpenLB release 1.8.1~\cite{olbRelease181} and is published open source under the GNU General Public License, version 2. 
Specific data is available upon reasonable request. 

\section*{Decleration of competing interests}
The authors declare that they have no known competing financial interests or personal relationships that could have appeared to influence the work reported in this paper.

\section*{Author contribution statement}
Conceptualization: MZ, SS; Methodology: MZ, SS; Software: MZ, DT, AK, MJK, SS; Validation: MZ, SS; Formal analysis: MZ, SS; Investigation: MZ, SS; Resources: MJK, MF; Data Curation: MZ, SS; Writing - Original Draft: MZ; Writing - Review \& Editing: AK, MJK, MF, SS; Visualization: MZ, SS; Supervision: MJK, MF, SS; Project administration: SS; Funding acquisition: MJK, MF, SS. 
All authors read and approved the final version of this paper.

\section*{Acknowledgement}
The authors gratefully acknowledge the computing time provided on the high-performance computer HoreKa by the National High-Performance Computing Center at KIT (NHR@KIT). 
This center is jointly supported by the Federal Ministry of Education and Research and the Ministry of Science, Research and the Arts of Baden-Württemberg, as part of the National High-Performance Computing (NHR) joint funding program (\url{https://www.nhr-verein.de/en/our-partners}). 
HoreKa is partly funded by the German Research Foundation (DFG).






\bibliographystyle{elsarticle-num-names} 
\bibliography{ref.bib}

\end{document}